\documentclass[twocolumn]{aastex62}

\usepackage{amsmath}
\usepackage{graphicx}

\usepackage{savesym}
\savesymbol{tablenum}
\usepackage{siunitx}

\restoresymbol{SIX}{tablenum}
\usepackage{xspace}

\DeclareSIUnit\parsec{pc}
\newcommand{\msun}{\textrm{M}_{\odot}}

\newcommand{\flash}{{\tt FLASH}\xspace}

\newcommand{\phfour}{{\tt ph4}\xspace}
\newcommand{\seba}{{\tt SeBa}\xspace}

\newcommand{\amuse}{{\tt AMUSE}\xspace}
\newcommand{\mul}{{\tt multiples}\xspace}

\global\def\bpath{./}
\global\def\ppath{./}

\received{December 21, 2018}
\submitjournal{ApJ}

\shorttitle{N-body---MHD}
\shortauthors{Wall et al.}

\begin{document}

\title{Collisional N-Body Dynamics Coupled to Self-Gravitating
  Magnetohydrodynamics Reveals Dynamical Binary Formation}



\author[0000-0003-2128-1932]{Joshua E. Wall}
\affiliation{Drexel University, Department of Physics and Astronomy,
  Disque Hall, 32 S 32nd St., Philadelphia, PA 19104, USA}
\correspondingauthor{Joshua Wall}
\email{joshua.e.wall@gmail.com}

\author[0000-0001-9104-9675]{Stephen L. W. McMillan}
\affiliation{Drexel University, Department of Physics and Astronomy,
  Disque Hall, 32 S 32nd St., Philadelphia, PA 19104, USA}

\author[0000-0003-0064-4060]{Mordecai-Mark Mac Low}
\affiliation{Department of Astrophysics, American Museum of Natural
  History, 79th St at Central Park West, New York, NY 10024, USA}
\affiliation{Center for Computational Astrophysics, Flatiron
  Institute, 162 Fifth Ave., New York, NY 10010, USA}

\author[0000-0002-0560-3172]{Ralf S. Klessen}
\affiliation{Heidelberg University, Center for Astronomy, Institute for Theoretical
  Astrophysics, Albert-Ueberle-Str.\ 2, 69120 Heidelberg, Germany }
\affiliation{Heidelberg University, Interdisciplinary Center for Scientific Computing, INF 205, 69120, Heidelberg, Germany}

\author[0000-0001-5839-0302]{Simon Portegies Zwart}
\affiliation{Leiden Observatory, Leiden University, Niels Bohrweg 2, 2333 Leiden, Netherlands}


\date{\today}

\begin{abstract}
 We describe a star cluster formation model that includes
 individual star formation
    from self-gravitating, magnetized gas, coupled to
collisional stellar dynamics. The model 
   uses the Astrophysical Multi-purpose Software Environment (\amuse) to
   integrate an adaptive-mesh magnetohydrodynamics 
code (\flash) with a fourth order Hermite N-body code (\phfour), a
stellar evolution code (\seba), and a method for resolving binary 
     evolution
(\mul). This combination yields unique star formation simulations
that allow us to study 
     binaries formed dynamically from interactions with both
     other stars and dense, magnetized gas subject to
stellar feedback during the birth and early evolution of stellar clusters. We
find that for massive stars, our simulations 
   are consistent with
the observed dynamical binary fractions and mass ratios. 
   However, our binary fraction drops well below observed values for
   lower mass stars, presumably due to unincluded binary formation during initial
   star formation. 
Further, we observe a build up of binaries near the hard-soft boundary that may be an important mechanism driving early cluster contraction.
\end{abstract}




\section{Introduction  \label{intro}}

The study of star cluster formation through simulations is a
non-linear physical problem with a wide range of scales. Clusters form
from turbulent, magnetized molecular clouds that are parsecs across,
yet the individual star formation process happens at scales of a
single AU or less \citep{Mac_Low_Star_Formation}. Further complicating
the issue, star formation contains a complex feedback loop in which
stars forming in one epoch affect proximal regions of current and
future star formation through radiation, winds and supernova
feedback. The gravitational contraction of molecular clouds, star
formation, stellar evolution, dynamical binary formation, and cluster
assembly and virialization all take place on timescales of 1--10
Myr. Resolving the relevant physical processes on all size and time
scales is computationally challenging.  As a result, approximations
for star formation are used that include sink particles representing
entire clusters \citep{Gatto_Walch_SILCC3_2016}, simplified stellar
feedback \citep{Dale_Winds_and_H2}, or softened gravitational dynamics
for stars \citep{Federrath_Sink_Particles}, or simulations neglect
important dynamical agents such as magnetic fields
\citep{rosen_unstable_2016} in order to make the problem tractable. 

In this study we describe numerical methods to resolve the dynamics of the stars and gas in order to study the formation of star clusters from gas collapse.
This includes coupling of the magnetohydrodynamics (MHD) code {\flash}
\citep{FLASH}, the N-body code {\phfour} \citep{ph4}, and the stellar
evolution code {\seba} \citep{Portegies_SeBa} using the Astrophysical
MUlti-purpose Software Environment \citep[{\amuse}][]{AMUSE}, 
and implementation of a subgrid model for the formation of
individual stars from sink particles. Since we focus on 
cluster formation and evolution as opposed to individual star formation,
we have chosen to use the initial mass function (IMF) as an input rather than a result of our simulations. To accomplish this, we sample a Kroupa IMF \citep{kroupa_IMF_2001} using a Poisson process,but still individually form each star in a way that conserves mass both locally and globally.

The natural environment to develop these methods is {\amuse}, as the
original intention in the development of {\amuse} was to allow for the
coupling of different astrophysical codes for multiphysics simulations
\citep{zwart_multi-physics_2013}. Further, multiple N-body and stellar
evolution codes already exist in {\amuse}, allowing us to change
methods as needed.  For example we could switch between {\seba} or
{\tt MESA} \citep{MESA_2011ApJS..192....3P} for stellar evolution
depending on the level of detail desired and computational cost
acceptable. This allows us to represent the stars in {\flash},
{\phfour} and {\seba} as a single data structure that can be modified
by any of the above codes, followed by propagation of the updated
information to all other running codes.
Interfacing {\flash} into the {\amuse} environment allows us to couple
the gravitational potentials computed by {\flash} and {\phfour} using a gravity bridge (see Sect.\ \ref{bridge}) directly using code in Python without major rewrites of either code.

In addition to interfacing {\flash} with {\amuse}, we have also made several additions to {\flash} itself. For the heating and
cooling of the gas we have modified the methods for atomic heating
and cooling of \citet{Joung_SN_driven_turb} and
\citet{ibanez-mejia_gravitational_2015} with the molecular and dust
cooling methods of \citet{Seifried_2011MNRAS.417.1054S}, which
themselves are based on \citet{neufeld_thermal_1995}. To do this we
have added our own model of
heating from the photoelectric effect to dust using either the
calculations from \cite{wolfire_2003} or
\cite{weingartner_photoelectric_2001}, which can be chosen
with a parameter switch. Finally, to solve for the both the degreee of ionization
and temperature of the gas as well as the dust temperature we have
implemented our own integrators based on well
known methods. 

We reserve a detailed examination of these
modifications to a subsequent paper,
where we will also detail modifications 
we have made to include feedback from individual stars. 
In this work, we focus on the coupling of gravity between {\flash}
and {\phfour} using a gravity bridge. 

In Sect.~\ref{bridge} we explain the concept of a gravity bridge and
how we implement it, while in Sect.~\ref{verification} we verify our
implementation. In Sect.~\ref{SF} we describe our method for
introducing star particles in regions of high gas density, and for
handling binary or higher-order systems in Sect.~\ref{multiples}.  We
define a demonstration problem in Sect.~\ref{demonstration}, and describe dynamical
binary formation occurring in our models in Sect.~\ref{binaries}.
Finally, we summarize our results in Sect.~\ref{conclusions}.

\section{Gravity Bridge} \label{bridge}

     \subsection{Implementation}
       \label{implementation}
Central to our implementation is the requirement to have fully collisional N-body
dynamics calculated for stars evolving in gas-rich regions.
To allow for physical interaction between the gas in {\flash} and the
stars in an N-body code, we implement a gravity bridge
\citep{Fujii_Makino_bridge} between the two codes. The method is
a ``kick-drift-kick,'' leapfrog-type integration
scheme with roots in the symplectic map method used by
 \cite{Wisdom_Holman_1991} to integrate the motions of
planets in the solar system. In \citet{Wisdom_Holman_1991}, the planets followed
an analytic Kepler orbit around the Sun while being perturbed perodically by each
other's gravitational acceleration. The scheme was 
extended by \citet{Fujii_Makino_bridge} to integrate a star cluster
subject to tidal effects inside a parent galaxy. While the method has
previously been used to couple gas in smoothed particle hydrodynamics
 (SPH) to stars contained in an N-body code
\citep[e.g.][]{Pelupessy_embedded_SC}, we have for the first time implemented
this method with an Eulerian, adaptive mesh refinement (AMR)
grid code. 
Here we briefly describe 
the {\amuse} bridge method, following \cite{Fujii_Makino_bridge}.

If we define the equation of evolution for our solution $f(q(t),p(t);t)$
in terms of the Poisson bracket
\begin{equation}
\frac{df}{dt} = \{f,H\},
\label{evol}
\end{equation}
where $H$ is the Hamiltonian of the system, and define an evolution operator $D_H$
\begin{equation}
  D_H = \frac{d}{dt} = \{\ \cdot \ ,H\},
  \label{Eq:Poisson_bracket}
\end{equation}
the formal solution for $f(t)$ is
\begin{equation}
  f(t+\Delta t) = e^{\Delta t D_H}f(t).
  \label{Eq:formal_solution}
\end{equation}

\citet{yoshida_construction_1990} noted that if $H$ (and therefore
$D_H$) is separated into kinetic and potential energy terms,
$H=K(p)+U(q)$ (with coordinates $q$ and momenta $p$), and $D_K$ and
$D_U$ are defined as in Eq. \ref{Eq:Poisson_bracket}, then the
exponential in Eq.\,\ref{Eq:formal_solution} can be approximated as
\begin{eqnarray}
  e^{\Delta t D_H} &=& e^{\Delta t \left( D_K + D_U\right)} \\
		 &=& e^{\Delta t} \prod\limits_{k=1}^{l}
		        e^{a_k D_K} e^{b_k D_U} + O(\Delta t^n),
  \label{Eq:Yoshida1990}
\end{eqnarray}
for suitable $l$, $n$ and constants $a_k, b_k$. In the simplest case,
$l = 2, n = 2$, $a_1=a_2=1/2$, $b_1 = 1$, and $b_2 = 0$, and the total
evolution of $f(t)$ becomes a second-order integration scheme upon Taylor expansion
\begin{eqnarray}
  f(t+\Delta t) &=& e^{\frac{\Delta t}{2}D_K} e^{\Delta t D_U}
  			e^{\frac{\Delta t}{2}D_K} f(t) \\
		 &=& \left(1+\frac{\Delta t}{2} D_K\right)
                        \left(1+ \Delta t D_U\right)
                        \left(1+\frac{\Delta t}{2} D_K\right) f(t).
                        \nonumber\\
  \label{Eq:Leapfrog}                        
\end{eqnarray}
We immediately recover the familiar kick-drift-kick formulation of the
leapfrog integrator, since
\begin{eqnarray}
  D_U q_i &=& \{q_i, H_U\} = \frac{p_i}{m_i} = \dot{q_i} =v_i \label{op:1} \\
  D_K p_i &=& \{p_i, H_K\} = - m_i \nabla V_K = F_g = m_i a_g, \label{op:3}
\end{eqnarray}
and the evolution of the system reduces to
\begin{eqnarray}
  v_i' &=& v_i + a_i(x) \frac{1}{2} \Delta t \\
  x_i' &=& x_i + v_i' \Delta t \\
  v_i' &=& v_i' + a_i(x') \frac{1}{2} \Delta t.
\end{eqnarray}

\citet{Wisdom_Holman_1991} noted that the Hamiltonian of a system
comprising two or more coupled subsystems can alternatively be split
into a set of secular evolution terms describing the internal
evolution of each subsystem and perturbation terms consisting of delta
functions, representing the interactions between the subsystems.
Following \citet{Wisdom_Holman_1991} and \citet{Fujii_Makino_bridge},
we split our Hamiltonian for each system into a sum of terms, $D_K$
and $D_D$, representing, respectively, the kick due to the external
perturbation and the drift due to internal (unperturbed)
evolution. Regardless of the split, the
\citet{yoshida_construction_1990} decomposition
(Eq.\,\ref{Eq:Yoshida1990}) still applies, and the evolution of the
system can be described by a scheme of the same form as
Eq.\,\ref{Eq:Leapfrog}.

\begin{figure}
  \includegraphics[width=\linewidth]{\ppath 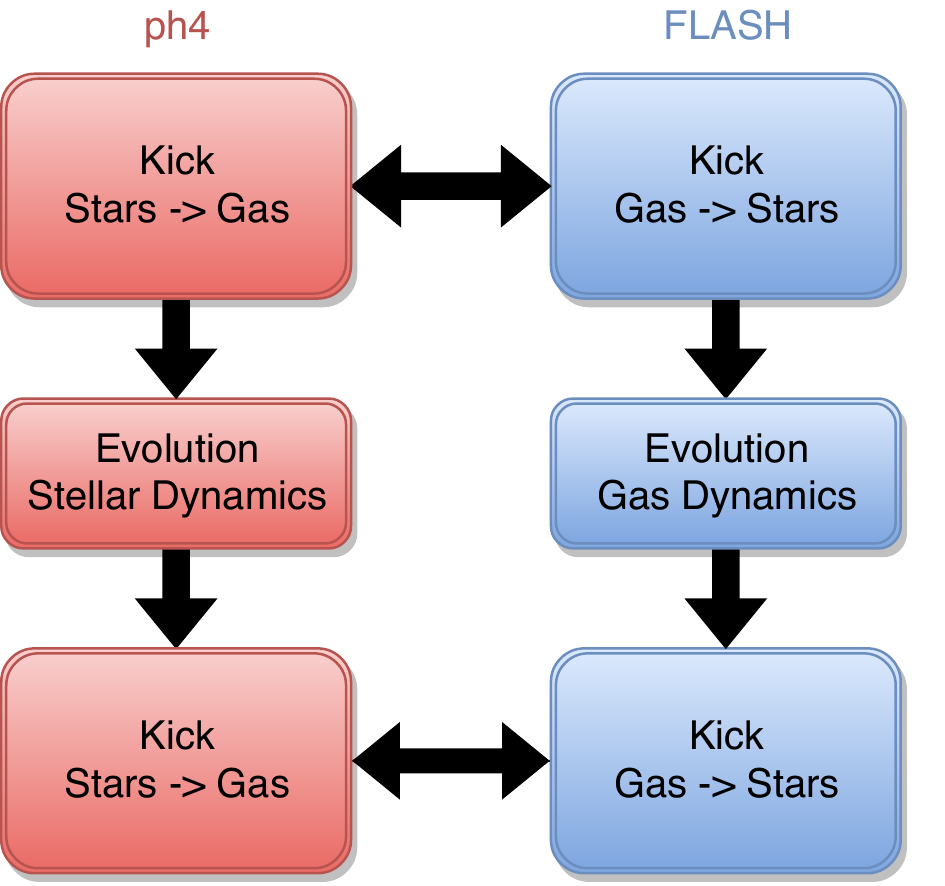}
  \caption{The bridge scheme implemented in {\amuse} using {\flash}
    for hydrodynamics, {\phfour} for N-body, and {\seba} for stellar
    evolution.
\label{BridgeFig.}}
\end{figure}

In our simulations, the subsystems are the stars (modeled using
{\phfour}) and the gas (modeled using {\flash}), so $D_K$ is computed
for the stars using the gravitational acceleration due to the gas, and
for the gas using the gravitational acceleration due to the stars.
For the drift operators, instead of deriving the drift from the
Hamiltonian as was done in Eq.\,\ref{op:1}, we 
use each subsystem's internal integration scheme 
as shown in Figure~\ref{BridgeFig.}. 
This means we now also
introduce any error from the internal schemes (fourth-order for
{\phfour} and second-order for {\flash}) to the formally symplectic
integration of the bridge, but we gain the ability to couple the codes
gravitationally. This error is found to be generally small, even for
fairly large bridge timesteps (see Sect.~\ref{verification}).
Our integration scheme for stars is
\noindent\begin{align}
&\mbox{kick}	&v'_{s,0} &=
     			v_{s,0} + \frac{\Delta t}{2}a_{g \rightarrow s,0} \\
&\mbox{drift} 	&x_{s,1}, v'_{s,1} &= {\phfour}
   			\left(x_{s,0}, v'_{s,0}, \Delta t\right) \\
&\mbox{kick}	&v_{s,1} &=
                         v'_{s,1} + \frac{\Delta t}{2}a_{g \rightarrow s,1}
\end{align}
where $a_{g \rightarrow s}$ is the gravitational acceleration on the
stars due to the gas. The stars receive an initial velocity kick from
the gas, then drift alone, then get a final velocity kick from the
gas.  The same considerations lead to a similar procedure for the gas:
\begin{eqnarray}
	v'_{g,0} &~=~& v_{g,0} + \frac{\Delta t}{2}a_{s \rightarrow g,0} \\
	x_{g,1}, v'_{g,1} &~=~& {\flash}
        		\left( x_{g,0}, v'_{g,0}, \Delta t \right) \\
	v_{g,1} &~=~& v'_{s,1} + \frac{\Delta t}{2}a_{s \rightarrow g,1},
\end{eqnarray}
where $a_{s \rightarrow g}$ is the gravitational acceleration on the
gas due to the stars.

At each bridge step, the gravitational acceleration due to gas in each
cell of the MHD code on the stars $a_{g \rightarrow s}$ is calculated
at the locations of each star in {\phfour}, and the gravitational
acceleration of each star on the gas $a_{s \rightarrow g}$ is
calculated at each cell site in {\flash}. For obtaining the
gravitational acceleration of the gas in \flash, we use the first
order finite differences of the potential calculated by the \flash
multigrid solver \citep{Ricker_MG_solver_2008}. For the acceleration
of the particles on the gas, we initially used the acceleration
directly from \phfour. However testing showed that combining two
different methods for stars on gas and gas on stars led to violations
of Newton's Third Law. We have therefore included a cloud in cell
mapping of the stellar masses onto the grid itself followed by the
same multigrid potential and acceleration solution method, allowing us
to use the same solver in both bridge directions to properly conserve
momentum during the interactions.

The method averages the gravitational acceleration of one code on the
other over the bridge time step, so the error in the bridge 
depends on the time step
$\Delta t$. Testing with different timesteps has shown
that $\Delta t \sim t_{\rm{ff}} /100$ is accurate enough to pass the
tests presented in the next section, 
where $t_{\rm{ff}}$ is the minimum free-fall time in the simulation,
although in practice we set
bridge timesteps much smaller than this, generally $\sim 2.5 \Delta
t_h$, where $\Delta t_h$ is the hydro time step. 
Runs at this time step still mean that
each code is taking numerous steps independently, making the whole
simulation more efficient overall. Also, since the two codes drift
independently, they can be in principle evolved in parallel for
another improvement in speed.

\subsection{Verification} \label{verification}

To test the gravity bridge we perform the test 
used by \citet{Federrath_Sink_Particles} for
sink particles when they were first incorporated into
{\flash}.  This consists of embedding
three particles at different radii on circular orbits centered on a
cloud of gas. The gas does not evolve and acts as a static potential,
This tests the actual interaction between gas and particles, 
unlike imposing a background static potential without 
representation on the grid. The density of the gas varies as
\begin{equation}
  \rho(r) = \rho(r_o) (r_o/r)^2,
\end{equation}
    where $\rho(r_o) = 3.82 \times \SI{e-18}{g.cm^{-3}}$ and $r_o = \SI{5e16}{cm}$,
which implies a gas mass of roughly \SI{3}{\msun} inside $r_o$. The three particles 
are placed at distances of \SIlist{e16;2e16;3e16}{cm} from the center of the gas 
cloud and have masses \SI{e-10}{\msun} such that the inter-particle gravity is very 
small compared to that of the gas. Each particle starts with a translational velocity 
for circular orbits
\begin{equation}
         v= (GM(r) / r)^{1/2}        =  \SI{895}{m.s^{-1}},
\end{equation}
which we lower by 2.3, 1.1, and 0.8\% respectively to account for the
non-singular nature of gravity on the grid at the origin
\citep{Federrath_Sink_Particles}.


\begin{figure}[h!]
	\includegraphics[width=\linewidth]{\ppath 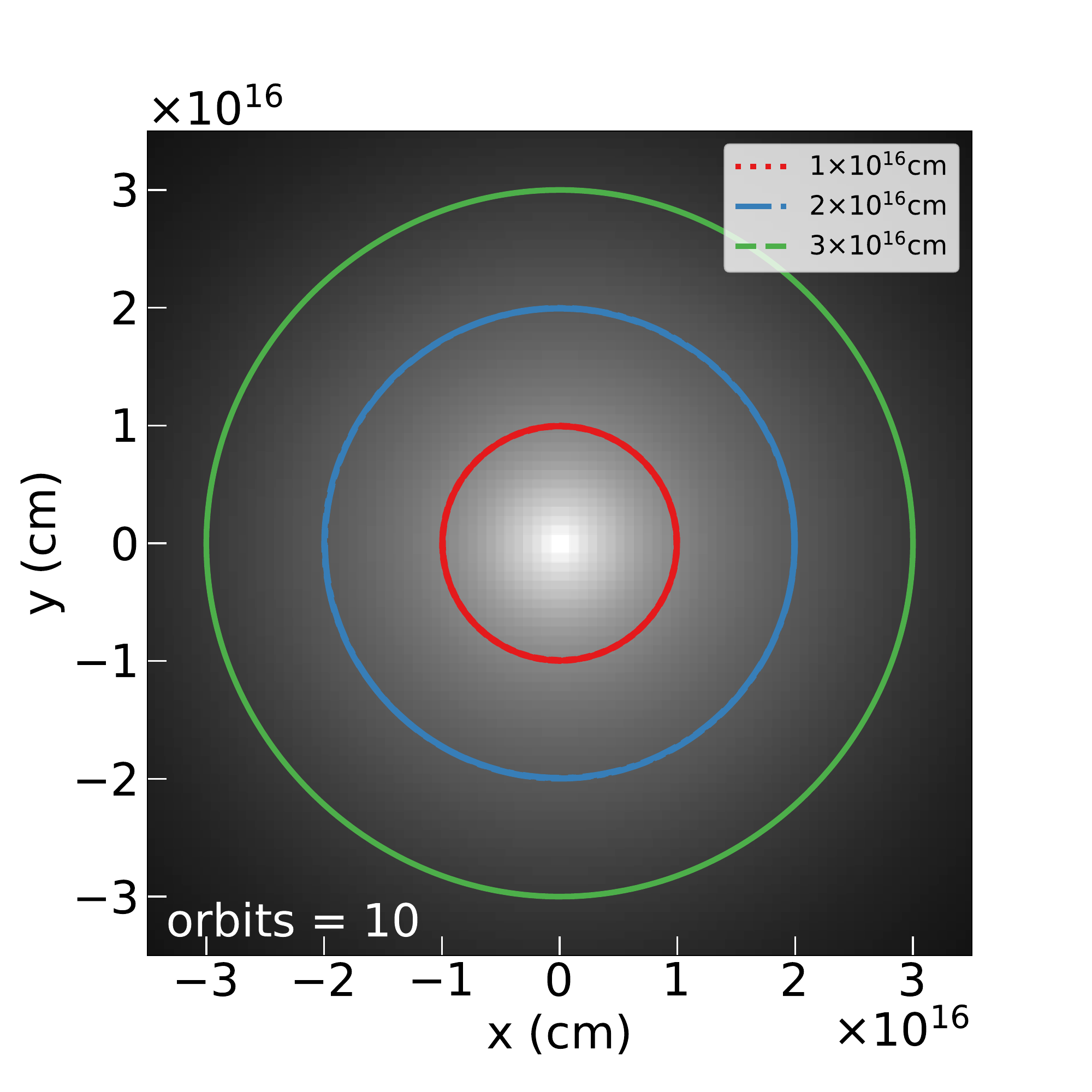}
	\caption{Orbital paths of three test particles after 10 orbits in an isothermal density profile where the gravitational acceleration of the particles is due to the bridge.
		\label{orbitsFig}}
\end{figure}

\begin{figure}[h!]
	\includegraphics[width=\linewidth]{\ppath 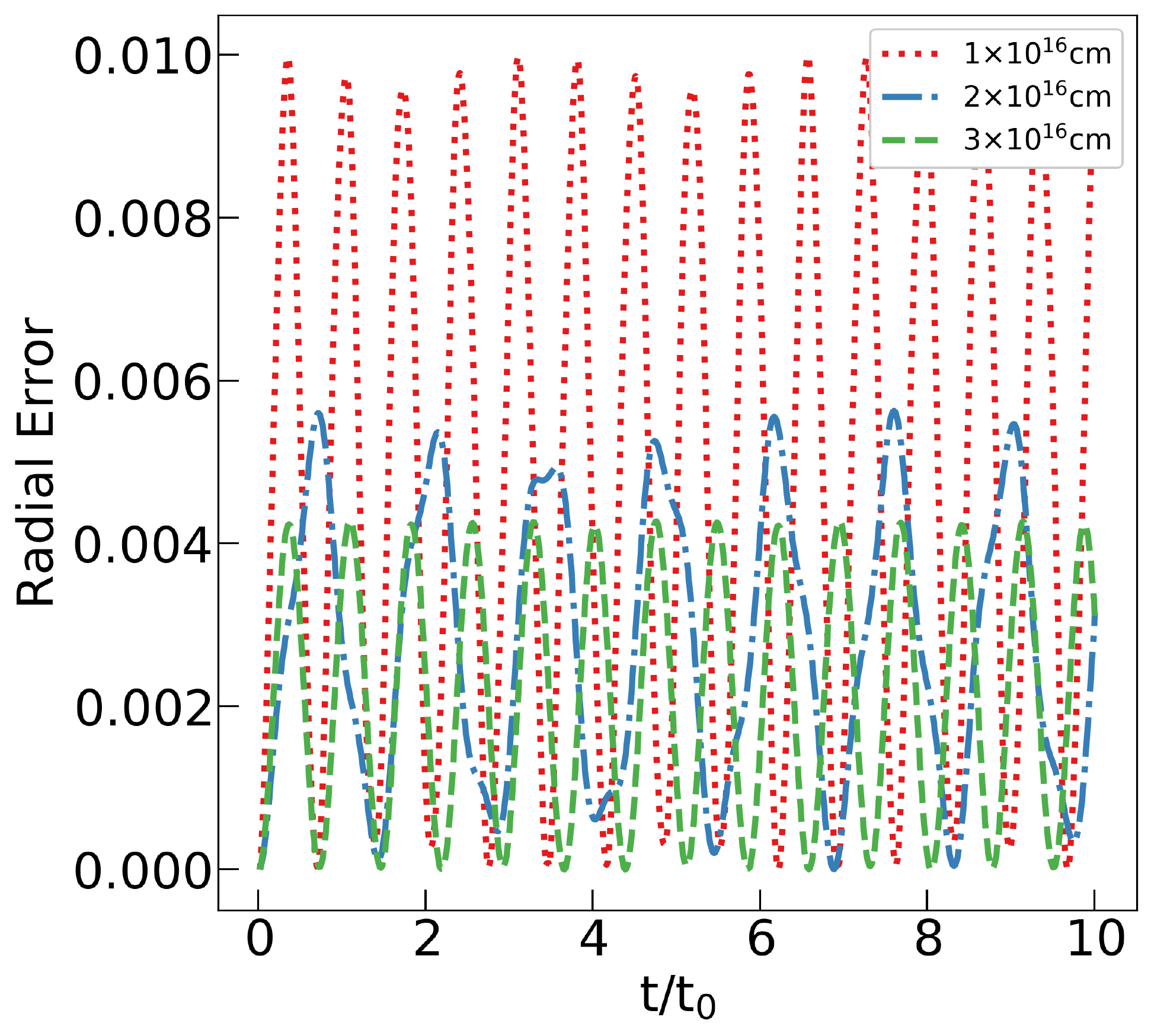}
	\caption{The fractional absolute error in 
		radius of the three test particles.
		\label{r_errFig}}
\end{figure}

\begin{figure}[h!]
	\includegraphics[width=\linewidth]{\ppath 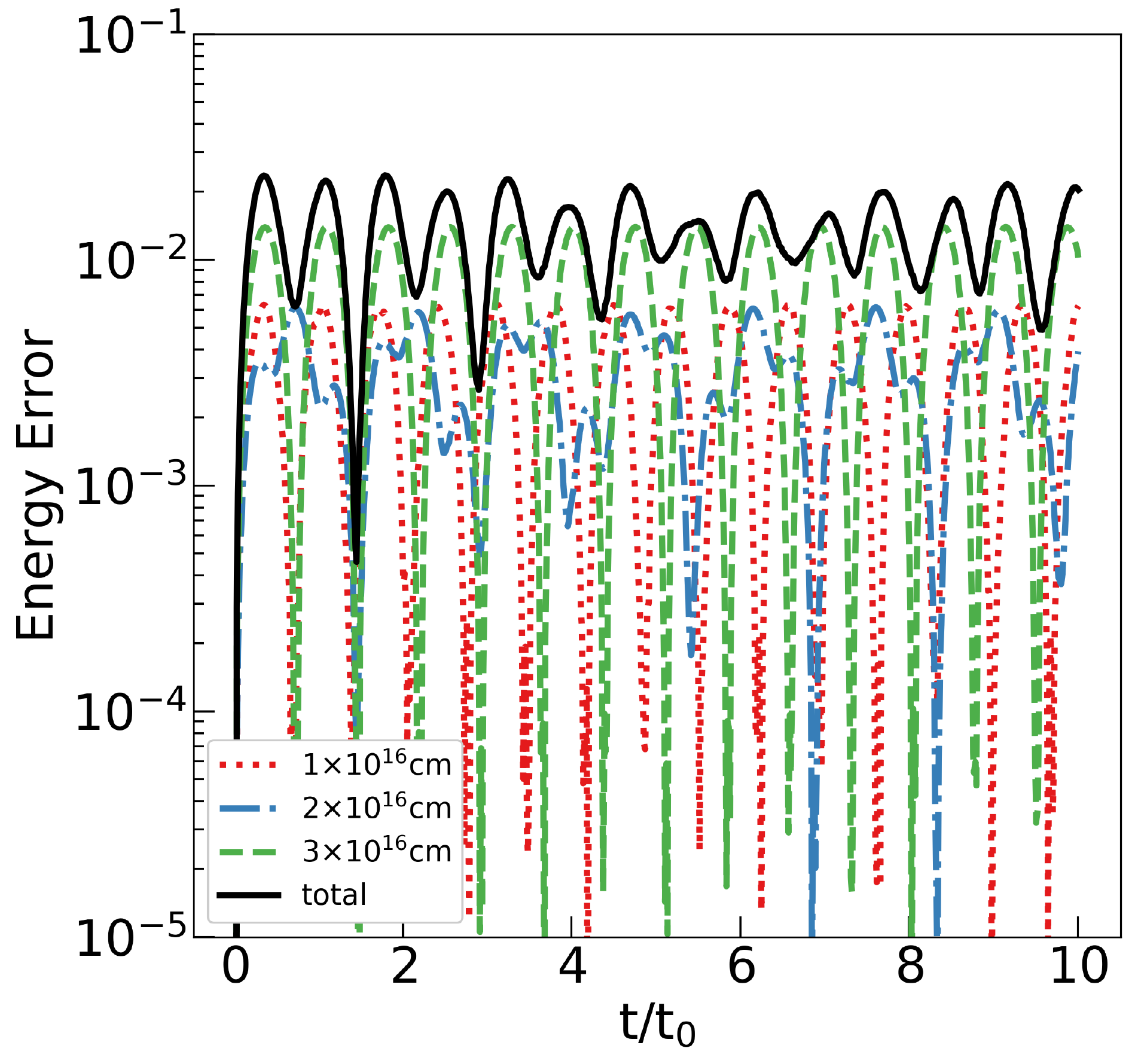}
	\caption{The fractional energy error of the three test particles. \label{e_errFig}}
\end{figure}

To check energy conservation we integrated 10 orbits of the innermost
particle to compare against \citet{Federrath_Sink_Particles}, with the
result shown in Fig.~\ref{orbitsFig}. Our integration appears to close
the orbits as well as the integration in
\citet{Federrath_Sink_Particles}, which used a second order leapfrog
scheme. However their integration produced larger errors in the outermost
orbit, while ours shows the most error in the innermost
orbit. \cite{Federrath_Sink_Particles} attributed the error in the outer orbit to the finite
effects of the grid (deviations from spherical symmetry) at the edges
of the grid. Our model does not show this effect strongly as the density drops
smoothly to the edge of the computational domain, while
in the Federrath test the cloud has a sharp edge at $\sim$
\SI{4e15}{cm} where the density changes by three orders of magnitude.

Although the orbits in  Figure~\ref{orbitsFig} are well closed, they do oscillate slightly about the proper path. 
This is more clearly seen in a plot of the fractional radial error 
(Fig.~\ref{r_errFig}). The resulting energy error, shown in Fig.~\ref{e_errFig}, 
never rises above $\sim 2 \%$. The larger radius error in the inner orbit
corresponds to the larger angular distance covered by the inner
particle between kicks, which in this test case were delivered at
fixed time intervals $\sim$ \SI{10}{yr}. The expected stability of
symplectic integrators is evident, and the energy 
error does not grow noticeably with time.

\section{Star Formation} \label{SF}

Capturing the range of scales in simulations is one of the core
challenges to overcome in conducting studies of star
cluster formation and the ISM in general. In order to account for the effects
of the surrounding medium, including its large scale turbulence,
magnetic fields and feedback, simulation boxes need to have sizes of
tens to hundreds of parsecs. However in order to properly capture star
formation for stars as small as a solar mass, including binary star
formation, simulations need to be able to resolve the Jeans length
$\lambda_J$
\begin{eqnarray}
  \lambda_J &=& 
   (\pi c_s^2 / G \rho)^{1/2}
\end{eqnarray}
which is on the order of or below a single AU.
     (Recent work does suggest that
perhaps only hundreds of AU need be resolved;
\citealt{sadavoy_stahler_embedded_2017}.)

To resolve the Jeans length in pure Eulerian hydrodynamics $\lambda_J$
must be resolved by at least four grid cells
\citep{truelove_jeans_1997}, while in MHD 
    at least six cells per Jeans length are needed to resolve Alfv\'{e}n waves
\citep{heitsch_gravitational_2001},
   and as many as 32 cells per Jeans length would be needed to
properly resolve self-consistent formation of magnetic fields through
the microturbulent dynamo \citep{federrath_new_jeans_2011}. These
requirements generally set the physical scales of the simulation, with
the computational expense increasing with dynamical range.
 
Many authors overcome this difficulty when simulating star formation in
large clouds by adding so called sink particles that represent entire clusters,
essentially truncating the small scales. Clusters are created from
Jeans unstable gas that is collected in sink particles
\citep{Bate_Sinks_1995,Krumholz_Sinks,Federrath_Sink_Particles} on the grid
\citep{Dale_Winds_and_H2,Gatto_Walch_SILCC3_2016}. This method
requires taking random samples from the IMF, 
which in turn requires that enough gas be collected that the
high-mass end of the IMF can be sampled appropriately. This typically
means that around \SIrange{100}{150}{\msun} must be collected in a
sink particle, which once sampled for a cluster population becomes a
single point source for all of the cluster's feedback.

The second difficulty in modeling star formation comes from the
effects of feedback at the protostellar and pre-main sequence phases. 
During the protostellar disk phase, accretion luminosity reduces fragmentation 
\citep{krumholz2007,bate2009, peters2010b, peters2011}.  This 
luminosity, along with protostellar jet driving, is
expected to reduce the efficiency of envelope
accretion \citep{matzner2015}.
All of these will have an effect on the final main sequence star 
that results. Generally these effects are replaced by a local star
formation efficiency parameter 
usually on the order
of \SIrange{0.1}{0.5}{}, which represents the fraction of gas that
survived the accretion process from the stars initial outer gas
envelope \citep[e.g.][]{padoan_infall-driven_2014}.

Here we use an efficiency parameter to account for proto- and
pre-stellar effects, creating our stars as main sequence 
objects. But to actually
create the stars we take a different, and perhaps more historical,
approach compared to
recent simulations. Instead of sampling an IMF directly
after collecting mass, we choose instead to take a Poisson sampling of
the number of stars in a given mass bin in the IMF for any given star
forming region as proposed by \citet{Sormani_sinks_particles},
\begin{equation}
  P_i (n) =  e^{-\lambda_i} \lambda_i^{n_i}/n_i !, 
\end{equation}
where $\lambda_i =  f_i M / \left\langle m_i \right\rangle$, 
$M$ is the total mass for a specific sample, $\left\langle m
\right\rangle$ is the average mass in the $i^{th}$ bin for the total
range of the IMF sampled over, $f_i$ is the fraction of the total
mass in the $i^{th}$ bin for the IMF range, and $n$ is the number of
stars for a specific sampling for which the probability $P$ is
returned. 

The idea of Poisson sampling for mass values has been used before to choose from
the IMF \citep{elmegreen_IMF_Poisson_1997}. It has the added
mathematical benefit that even when sampling one star at a time the
sum of all the samples will always reproduce the parent sample, since the product of the subset Poisson distributions of $n_i, n_j$ with mean values $\lambda_i, \lambda_j$ is equal to the Poisson distribution of the entire set $N$ with the mean being $\lambda_i+\lambda_j$:
\begin{eqnarray}
P(N) &=& \sum_{i=0}^{N} P(n_i,\lambda_i | n_j, \lambda_j) \\
&=& \sum_{i=0}^{N} \frac{\lambda_i^{n_i}}{n_i!} e^{-\lambda_i} \
                   \frac{\lambda_j^{n_j}}{n_j!} e^{-\lambda_j} \\
&=& \frac{\left(\lambda_i+ \lambda_j\right)^N}{N!} e^{\lambda_i+\lambda_j},
\end{eqnarray}
from the binomial theorem.

On the face of it, it would seem that the same considerations of having
enough mass to sample each mass bin appropriately would apply to our
Poisson sampling as well \citep[see for
  example][]{Sormani_sinks_particles}, since even an input of
\SI{1}{\msun} of gas can result in an unphysical \SI{20}{\msun} star,
even if we lack enough mass in the simulation to create it. Suggestions
to overcome this difficulty in star formation methods have
been to violate local mass conservation by sampling from all sink
particles at once, or to simply sample all gas over a given density
threshold throughout the simulation \citep{fujii_zwart_IMF_2015}.

Instead we \textit{invert} this process. We use the sink particle
routines of \citet{Federrath_Sink_Particles} in {\flash} to identify
star forming regions. Every time a sink particle forms because a
region has become Jeans unstable, we create a list of stellar masses
for that sink particle by sampling the IMF with our Poisson process
with \SI{e4}{\msun} of stars created at once, before the sink accretes
any gas. The number of stars from the Poisson sampling in each mass
bin is returned, and then we randomly sample from 
the  \citet{kroupa_IMF_2001} IMF 
in each bin to give actual masses to every 
star, with the sampled mass bracketed between
\SIrange{0.08}{150}{\msun}.
    We choose a minimum mass of 0.08~M$_{\odot}$ for the IMF.
We then randomize the entire list of
stars created. Once the sink particle obtains enough mass to create
the first star in the list, this star is removed from the list and
placed into the simulation, after which the sink must then gather
enough mass for the second star in the list.

This method allows us to form particles star by star, without 
any violation of mass conservation. Each particle can take on the
local momentum, mass and velocity of the sink at the time of
formation. Also if a massive star forms, it has the chance to shut
down local (or non local) star formation in the simulation
by preventing further accretion, 
allowing the effects of feedback on star formation to be properly
analyzed. Furthermore, since stars are formed individually, 
gravitational interactions between stars and with the surrounding gas
can lead to binary formation and stellar ejections that can have
important dynamical effects on the clusters and their surrounding
natal gas clouds.

\begin{figure}[h!]
  \includegraphics[width=\linewidth]{\ppath 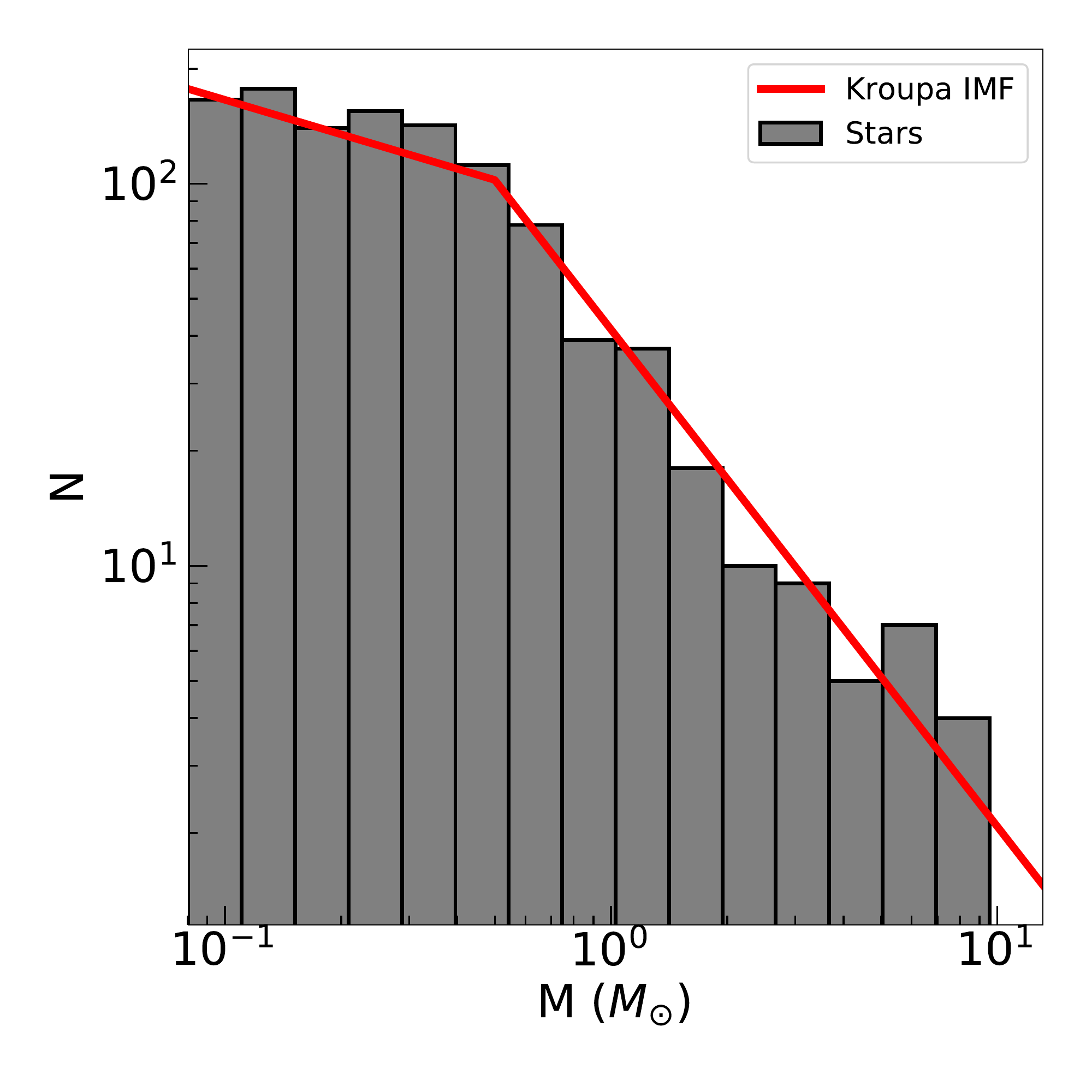}
  \caption{The mass function of stars from run
    M3f presented in the Results section. The Kroupa IMF is shown for comparison, 
    normalized to the same number of total stars as in the simulation, here 1100. 
    \label{fig:imf}}
\end{figure}

Each gas-gathering sink particle has an accretion 
radius of 2.5 times the smallest grid cell, to capture the local flow
for accretion of the gas \citep{Federrath_Sink_Particles}. Since this 
is the best resolved location we have for star formation, star 
particles are placed randomly within this radius using an isothermal 
spherical density profile \citep{binney2011galactic}. This allows some 
stars to form on the edges of these regions, but with smaller 
probability. 
We sample the velocity for the
star from a Gaussian profile centered on the sink velocity, using the
local gas sound speed as the variance. 
Fig.~\ref{fig:imf} is
the resulting mass distribution for a typical small cloud (\SI{e3}{\msun})
run using this method. 

\section{Multiple Stars}\label{multiples}
The formation of close binaries and higher-order systems can lead to
the effective time step
shrinking to a small fraction of the binary orbital period,
preventing further integration of the solution due to the 
extreme computational expense.
Normally, in N-body codes, the solution to this problem is to introduce a
specialized treatment of close encounters, through regularization of
the equations of motion \citep{Aarseth_KS_regularization_1974}
or some other approximate treatment of close encounters
\citep{Portegies_Zwart_runaway_colls_clusters_1999}. 
Several of the N-body
modules in AMUSE have the ability to incorporate such treatments, but
for a general and minimally intrusive solution within the \amuse\
framework, we prefer to handle close encounters using an external
module, as we now describe.

The basic simplification in the approach we use is that the N-body
code manages only the centers of mass of stable multiple systems.
These include binaries, a stable hierarchical triples
according to the \citet{mardling2008dynamical}
criterion, or a higher-order multiple systems in which
the Mardling criterion applied to the outermost orbits indicates
stability.  Close encounters are resolved using the {\mul} module
\citep{Mcmillan_Art_of_AMUSE}, which keeps track of the internal
structure of all multiple systems and manages interactions between
them.  To operate with this module, an N-body code must be modified to
detect close encounters and return immediately to the top-level \amuse\
script controlling the simulation, where appropriate means are taken
to resolve the encounter.  Such functionality is straightforward to
add, and is applied at the end of every N-body step.

In our case, the \phfour\ module checks for pairs of particles that
satisfy the stopping conditions that (1) they are approaching, (2)
they have separations less than twice the 
sum of their effective dynamical diameters
(a tunable parameter set at runtime to be $\SI{100}{AU}$
 for all stars and twice the semi-major
axis of a binary), and (3) they are relatively unperturbed by their
next nearest neighbor (with the ratio of accelerations 
$\gamma_p<0.02$ in the terminology discussed
below in Sect.~\ref{binaries} and represented by Eq.~\ref{pert_eqn}).
Once the stopping condition is triggered, any internal structure in
the two interacting particles is restored, and the entire system is
moved to a separate code designed for small-N encounters, aptly named
{\tt smallN} \citep{Hut_better_LF,McMillan_Hut_smallN}.  The {\tt
  smallN} code models the encounter as a scattering experiment,
terminating when the system has resolved itself into a collection of
mutually unbound single stars or stable multiples (as just defined).
The internal structure of the stable multiples is saved, their centers
of mass are placed back in the N-body code, and the integration
continues.  In this way, arbitrarily complex hierarchical
configurations can form and interact, and their dynamical histories
can easily be monitored.  This treatment of multiples is unusual in
the N-body community, but similar implementations are widely used in
Monte Carlo models of cluster dynamics
\citep{chatterjee2010,hypki2013}.

Currently, the internal structure of a multiple is simply frozen until
its next close encounter.  Secular internal evolution or perturbations
due to encounters too wide to trigger a stopping condition are not
included.  Binaries on wide or strongly perturbed orbits are not
merged into their center of mass; instead, their components are
returned directly to the N-body code.  We note that, although \phfour\
evolves only the centers of mass of multiple star systems, 
for all feedback and bridge calculations the 
individual component stars are used directly.

\section{Demonstration Problem} \label{demonstration}

As a demonstration of our method, we simulate star formation in
turbulent spheres of gas. For gas dynamics in \flash we use the unsplit MHD solver 
\citep{Lee_USM_2013} with third order PPM reconstruction 
\citep{Colella_and_Woodward_PPM_1984} and the HLLD Riemann solver 
\citep{Miyoshi_HLLD_solver}, while for solving Poisson's equation for gravity we use 
the multigrid solver of \cite{Ricker_MG_solver_2008}.  We include feedback in these 
runs from radiation, winds and supernovae that we will describe in a
subsequent paper, since the results we describe here are not
strongly affected by the feedback. We initialize the density
field with the commonly-used, initially spherically symmetric,
Gaussian, gas distribution of \cite{Bate_Sinks_1995},
while the velocity field is generated with a turbulent Kolmogorov
velocity spectrum \citep{WunschCloud} for the dense gas.  
All runs have eight levels of AMR refinement with the exception of M3f, which has seven, with refinement triggered by the Jean's criterion described in
\citet{Federrath_Sink_Particles}. All the runs except M3V2 have all
three stellar feedback methods (winds, radiation and supernovae)
switched on, although no run has yet to experience a supernova event.

\begin{figure*}
	\gridline{\fig{\ppath 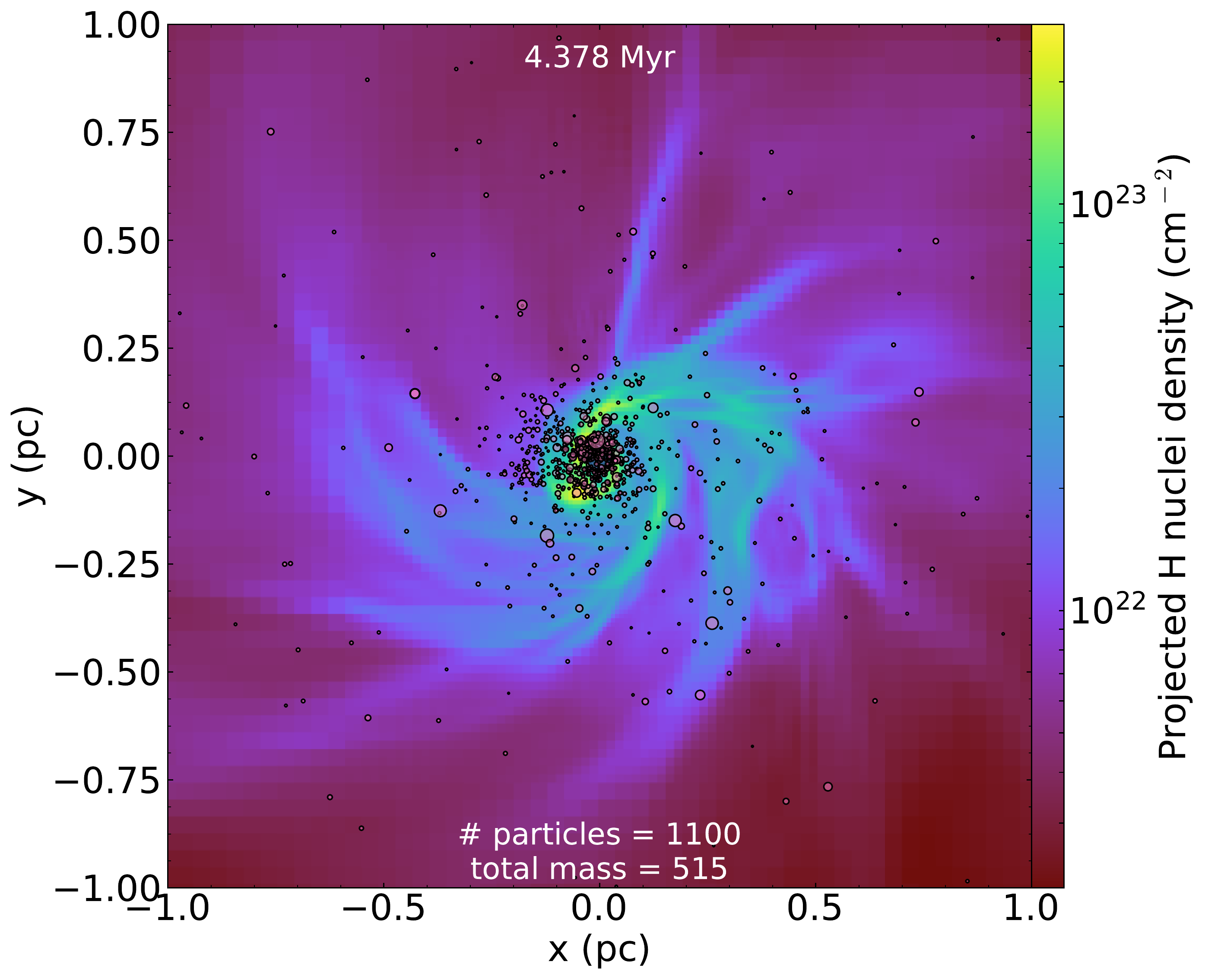}{0.5\textwidth}{(a)}
		      \fig{\ppath 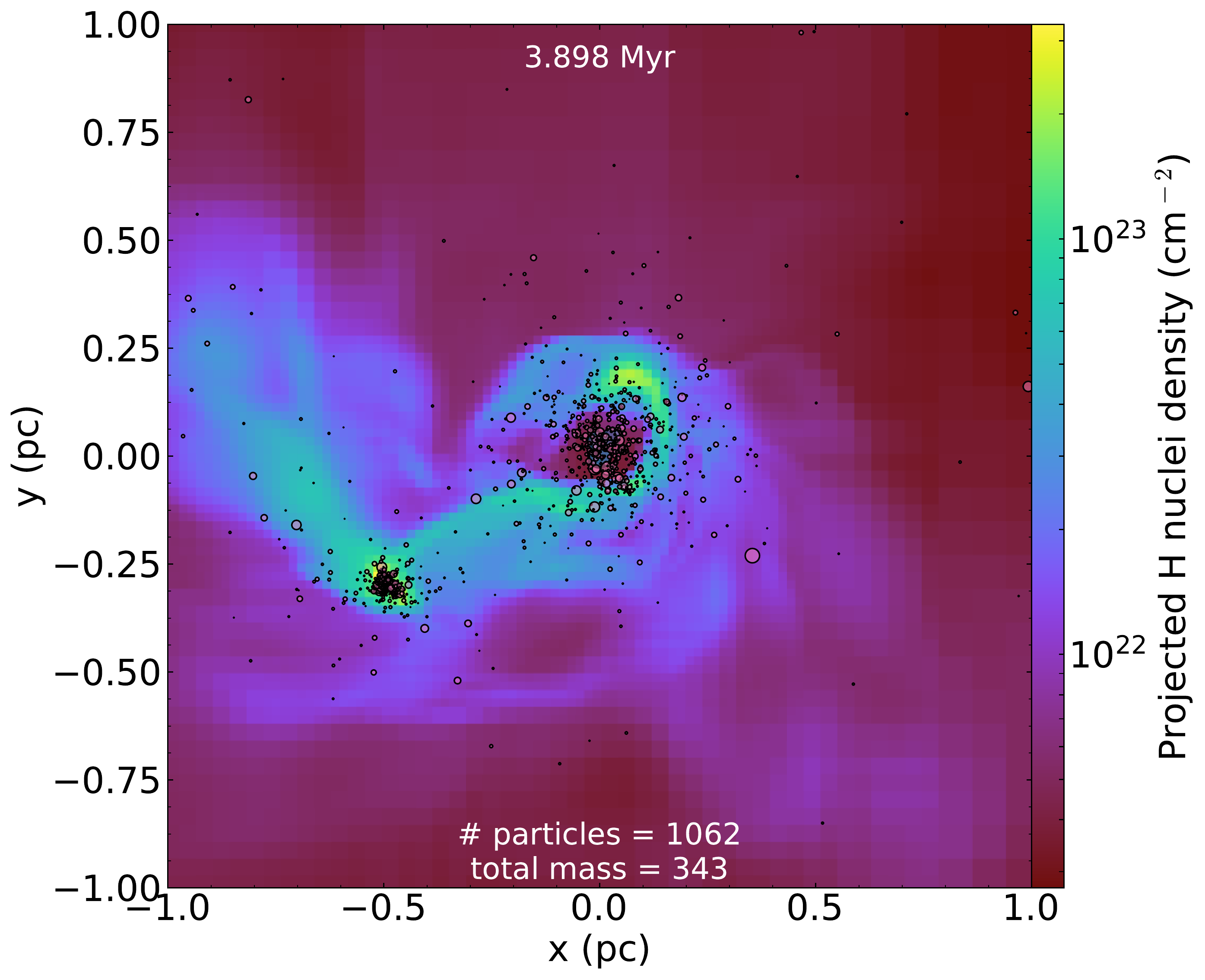}{0.5\textwidth}{(b)}
	}
	\gridline{\fig{\ppath 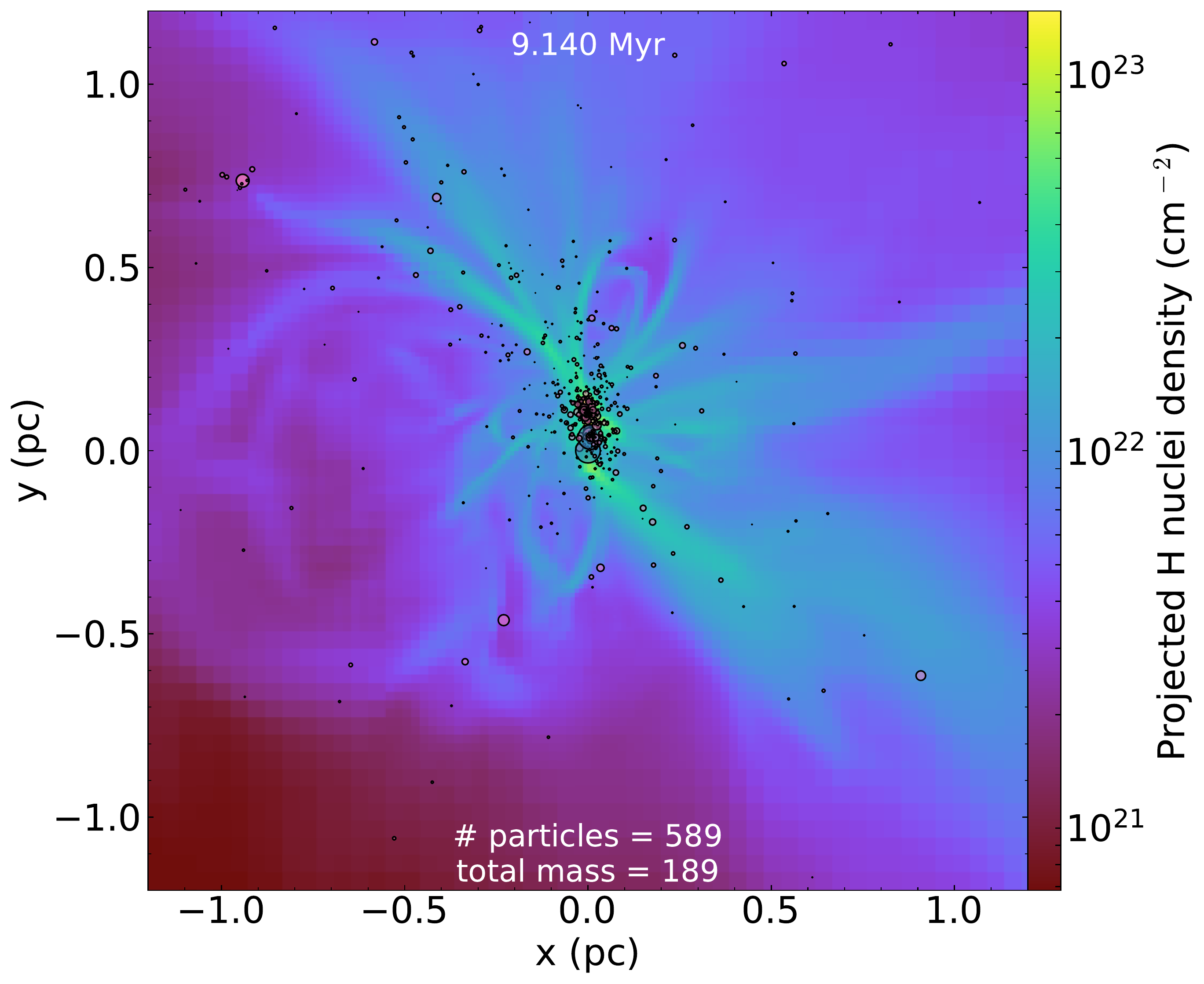}{0.5\textwidth}{(c)}
	          \fig{\ppath 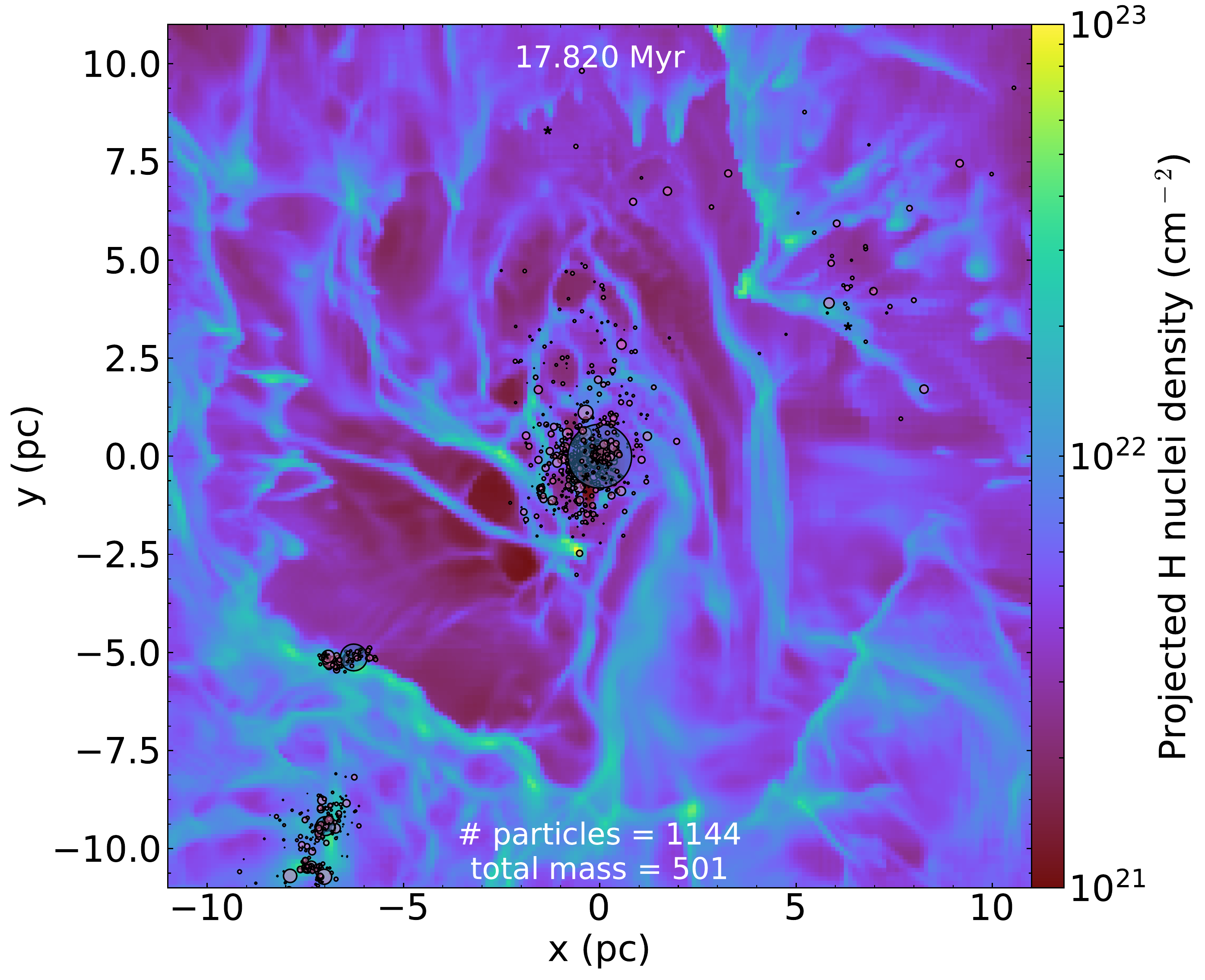}{0.5\textwidth}{(d)}
	}
	\caption{Projected number density along the $z$-axis for runs (a) M3 (b) M3f (c) M4f and (d) M5f at the last data file from each run. The area of the circles representing stars are
	proportional to their mass, while the locations of sink particles are
	shown by white star symbols. Feedback is most effective in run (b) where multiple feedback stars have sunk together to the center of the cluster and in (d) due to the $97~\msun$ star in the center of the image.\label{fig:ndens_grid}}
\end{figure*}

We use total masses of $M = $\SIlist[list-units=single]{e3;e4;e5}{\msun} and 
Gaussian density profiles with variance
$r_o =$ \SIlist[list-units=single]{5;10;50}{pc} respectively.
These length scales are chosen to roughly match
 the average density scales of clouds of these
masses \citep{stahler2008formation}. Note this means the larger clouds
have significantly longer free fall times. Outside of the sphere the
density is chosen to roughly match the ISM density for a containing medium
based on the size and density of the sphere itself (i.e. for the $10^3 \msun$ 
sphere, with higher density, the containing medium was assumed to be cold
neutral medium, while for the $10^5 \msun$ sphere the containing medium is warm and 
ionized). Then the temperatures are chosen to keep the sphere and containing medium
in pressure equilibrium.
The physical grid domain sizes $D$,
    listed in Tab.~\ref{tab:table},
    are $\sim$ \numrange{1.3}{1.5} times the Gaussian width $r_0$ in each case.
All models reported here were initialized with virial ratio 
   of kinetic to potential energy of 0.2.
We choose this initially low virial parameter to ensure quick cloud
collapse even before all of the turbulence decays within a free-fall time
\citep{Mac_Low_Kinetic_1998}.  As 
expected our spheres rapidly collapse into filamentary structures and
begin forming stars (see Fig.\ \ref{fig:ndens_grid}).

\begin{deluxetable*}{ccccccc}
\tablecaption{Parameters for each of the four runs described
  here including mass $M$, total number of stars $N_s$ at end of
  run $t_{\rm end}$ when analysis was performed, time
  at first star-forming event $t_{\rm sf}$, 
  the cell size $\Delta x$ at maximum refinement and the domain size D.
  Note that M3 and M3f used different random turbulent
  patterns initially, explaining their different values of $t_{\rm sf}$.
 \label{tab:table}}
\tablehead{
	\colhead{Run\tablenotemark{a}} & \colhead{$M$ ($\msun$)}
	 & \colhead{$N_{s}$} 
	 & \colhead{$t_{\rm sf}$ (Myr)} 
	 & \colhead{$t_{\rm end}$ (Myr)}
 	 & \colhead{$\Delta x$ (pc)}
	 & \colhead{D (pc)}}
\startdata
	M3     & $10^3$ & 1100 & 2.86  & 4.38  & 0.01 & 10 \\ 
	M3f    & $10^3$ & 1062 & 2.31  & 3.90  & 0.02 & 10 \\ 
	M4f    & $10^4$ & 589  & 7.76  & 9.14  & 0.05 & 12 \\ 
	M5f    & $10^5$ & 1144 & 15.38 & 17.82 & 0.2  & 110
\enddata
\tablenotetext{a}{Runs ending in ``f'' include radiation, winds, and
  supernovae.} 
\end{deluxetable*}

The four models we analyze are the current 
snapshots of our first runs, listed in
Tab.~\ref{tab:table} and shown in 
Fig.\ \ref{fig:ndens_grid}. Note although M5f 
is both more massive and significantly older, 
it also contains a $97 \msun$ star that is
shutting down star formation in a large volume. 
Therefore the number of stars is comparable to 
the much younger star forming regions in other runs.
The larger runs have larger minimum cell sizes, since all runs have
the same number of refinement levels, but they also have more
individual filaments and cluster forming regions. Finally we note that simulations
of this nature are in general highly stochastic and therefore only predictable in a statistical sense.

\begin{figure*}
	\gridline{\fig{\ppath 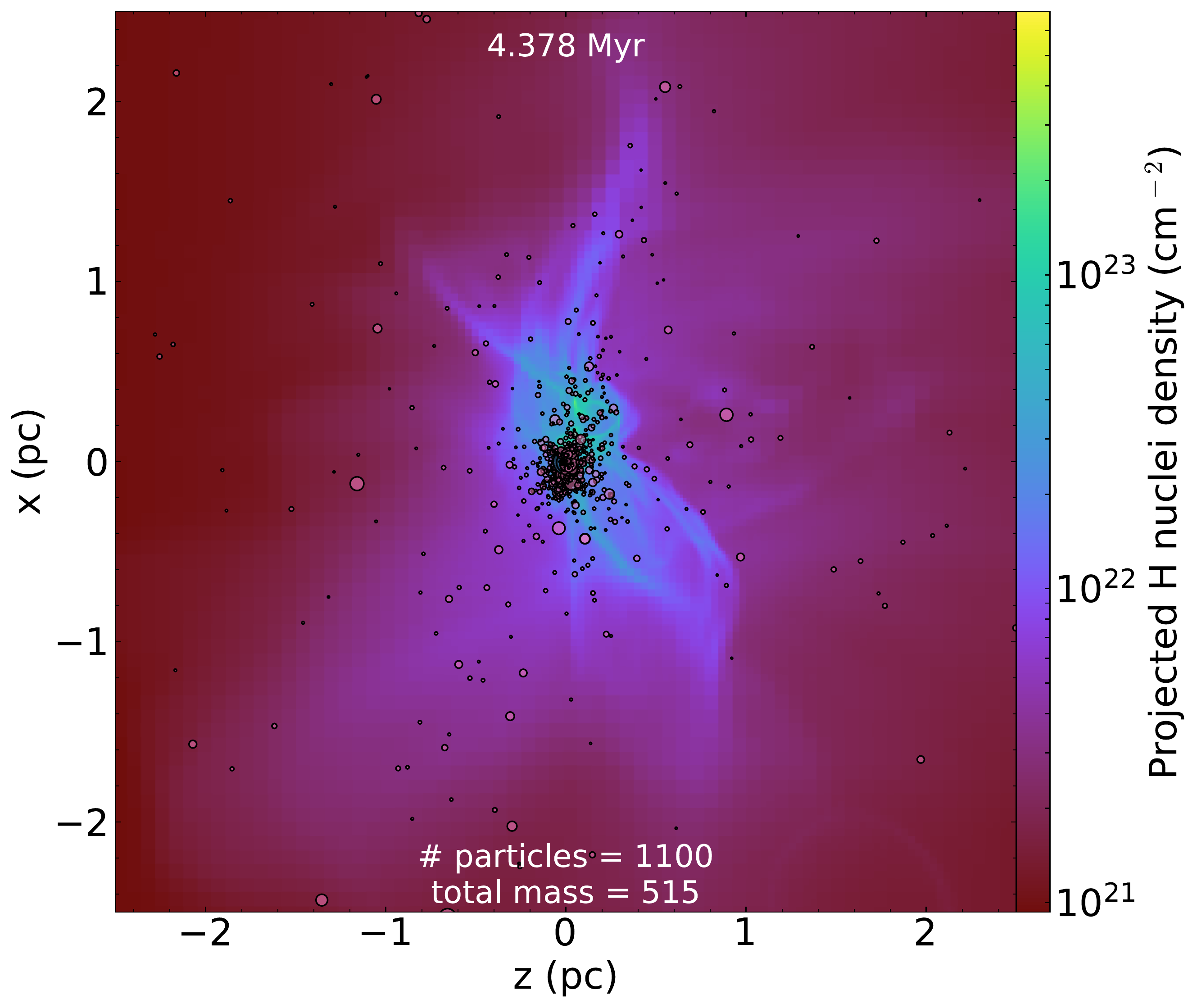}{0.5\textwidth}{(a)}
		\fig{\ppath 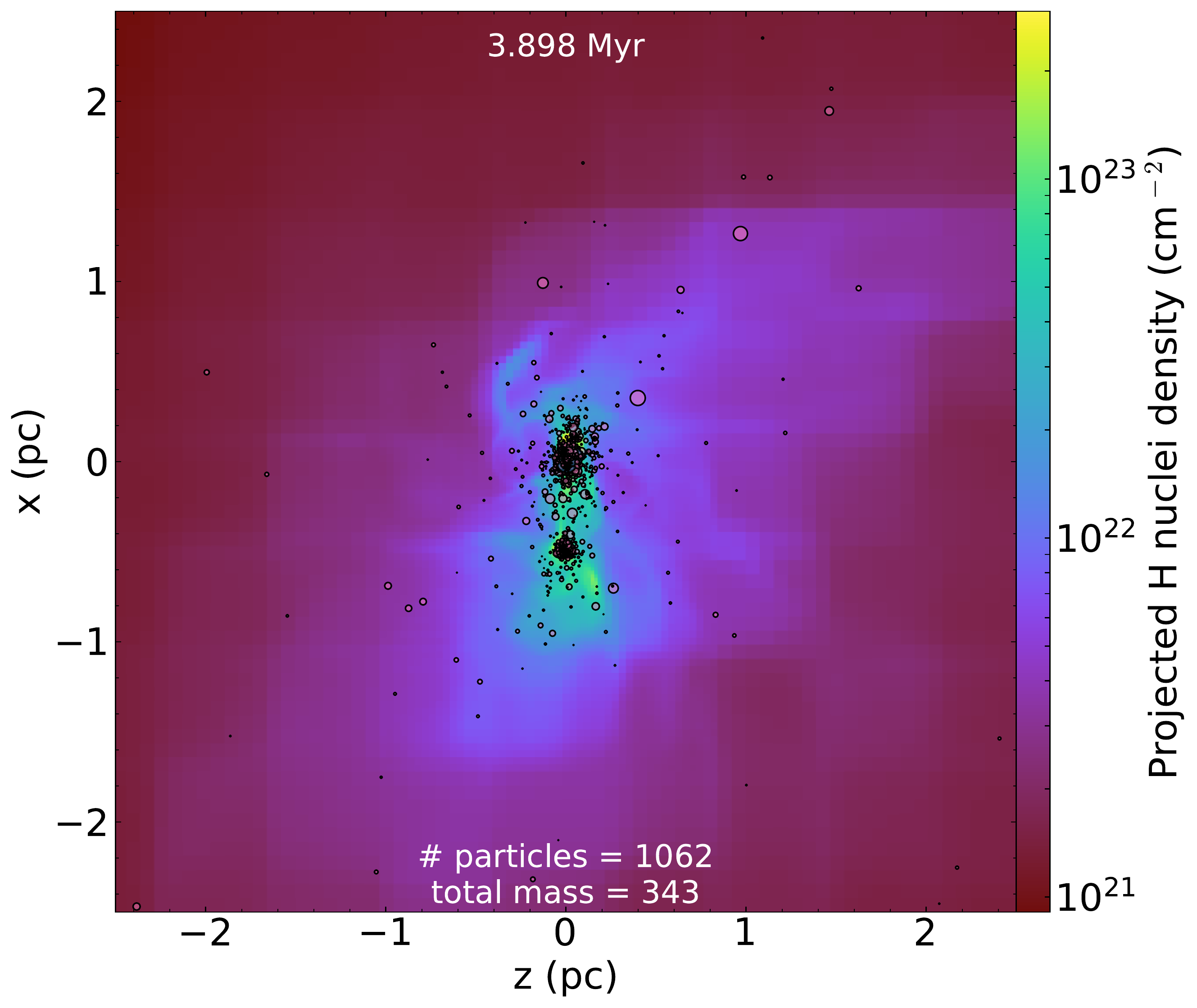}{0.5\textwidth}{(b)}
	}
	\gridline{\fig{\ppath 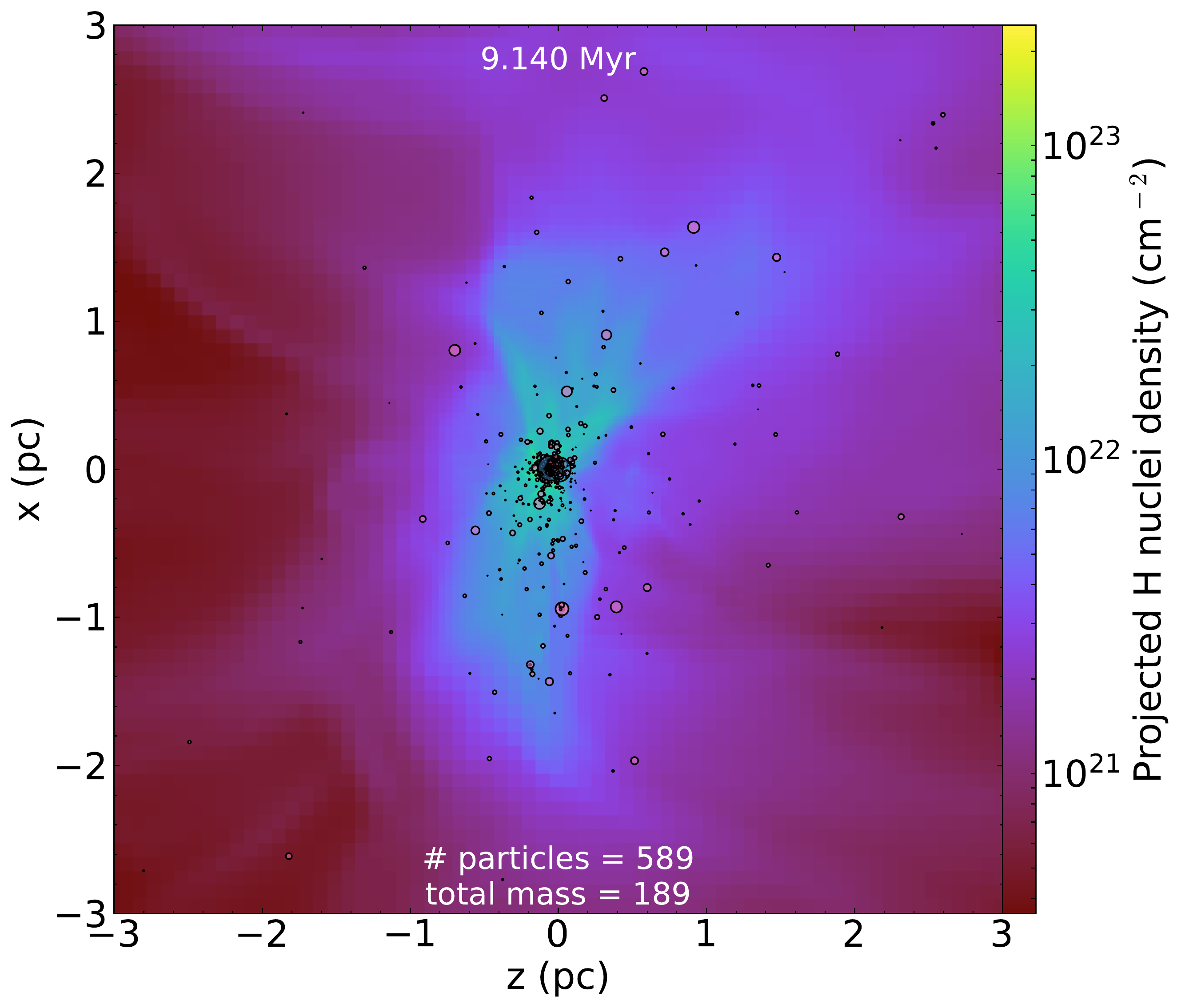}{0.5\textwidth}{(c)}
		\fig{\ppath 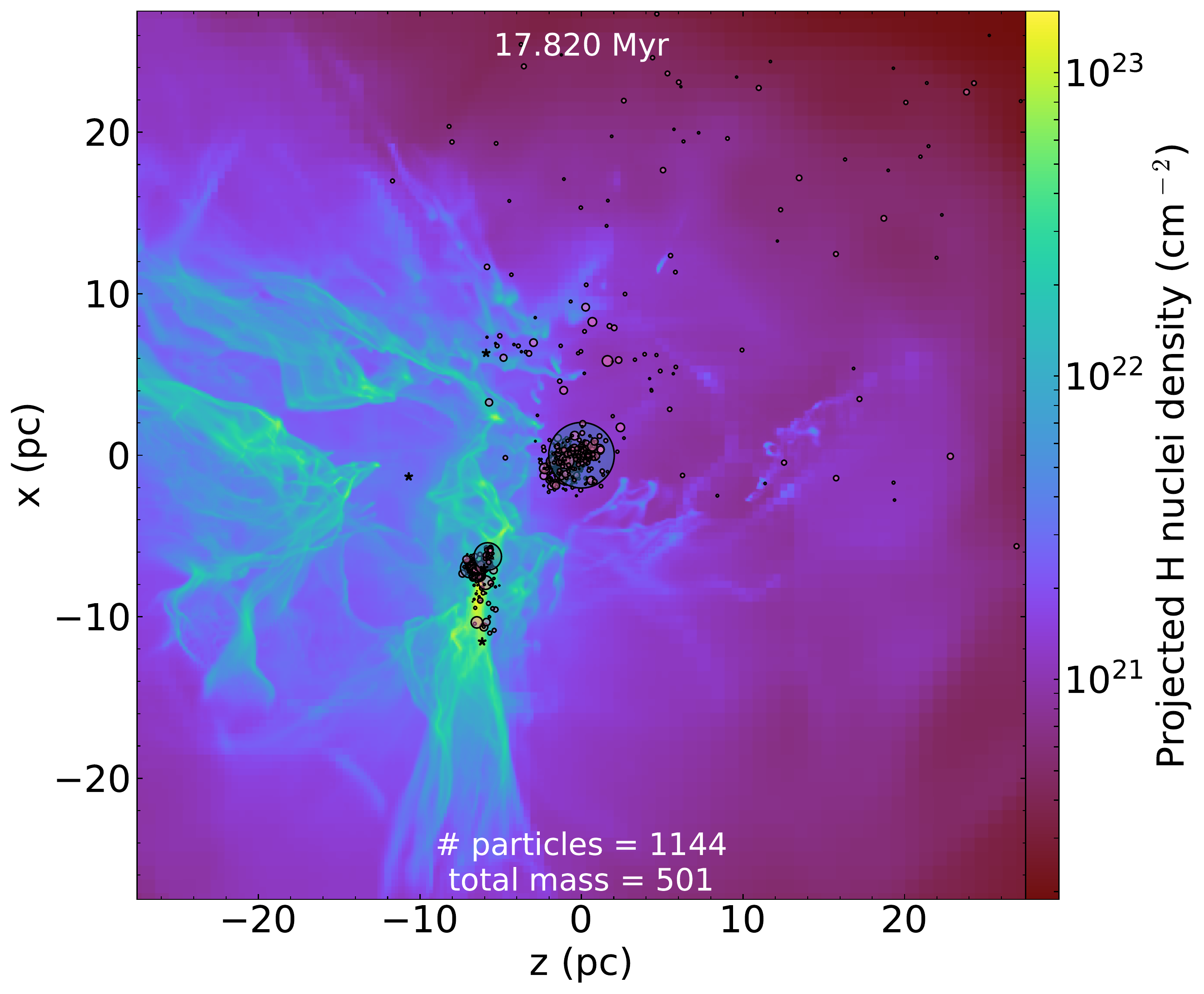}{0.5\textwidth}{(d)}
	}
	\caption{Projected number density along the $y$-axis for runs (a) M3 (b) M3f (c) M4f and (d) M5f at the last data file from each run. These images are zoomed out by a factor of $\sim 2$ compared to Fig.\ \ref{fig:ndens_grid} to better show the overall structure. (a) and (c) have fully collapsed and merged, while (b) is in the process of merging two subclusters, and (d) is still scattered. (d) also shows signs of partial disruption as the cloud was destroyed above and right of the $97~\msun$ star, causing many cluster members to become unbound. \label{fig:ndens_grid2}}
\end{figure*}

\section{Binaries}\label{binaries}

Given the collisional nature of our coupled code, 
the possibility exists of dynamical binary formation by
interactions between stars and with the gas.  Indeed, all
four runs examined here formed binaries. Note that here when
we refer to binaries, we mean any particles that are bound.  Not all
of these will necessarily be merged into root particles in
\mul (Sec. \ref{multiples}). The \mul\ code is a numerical solution to the
problem of prohibitively short timesteps, but will only act on
the tightest physical multiple systems.

To identify binaries in our simulations, we first calculate the
relative energies of all stars with respect to each other, keeping only those that
are bound to each other. We identify those that
are mutually bound, which is our initial list of binary candidates. We
then test each binary, consisting of stars with masses $m_1$ and $m_2$
and semi-major axis $a$, for perturbation by any star with mass $m_p$
and distance $d$ from the binary center of mass,
\begin{eqnarray}
\left| \frac{m_p m_1}{\left(d-a\right)^2}
    - \frac{m_p m_2}{\left(d+a\right)^2} \right|
    ~<~ \gamma_p \frac{m_1 m_2}{4a^2}, \label{pert_eqn}
\end{eqnarray}
where $\gamma_p$ is the chosen limiting ratio of 
accelerations. As a guide for choosing $\gamma_p$, we consider cases 
in which all three stars have equal mass. Then for ratios
$d/a =$ 2, 5, and 10, inverting Eqn. \ref{pert_eqn} gives
$\gamma_p \sim$ 3.0, 0.14, and 
0.016 respectively. For the preliminary analysis that we present here,
we chose $\gamma_p=3.0$ and combine the results of all four
runs, which yields 85 binaries.

\begin{figure}[h!]
  \includegraphics[width=\linewidth]{\bpath 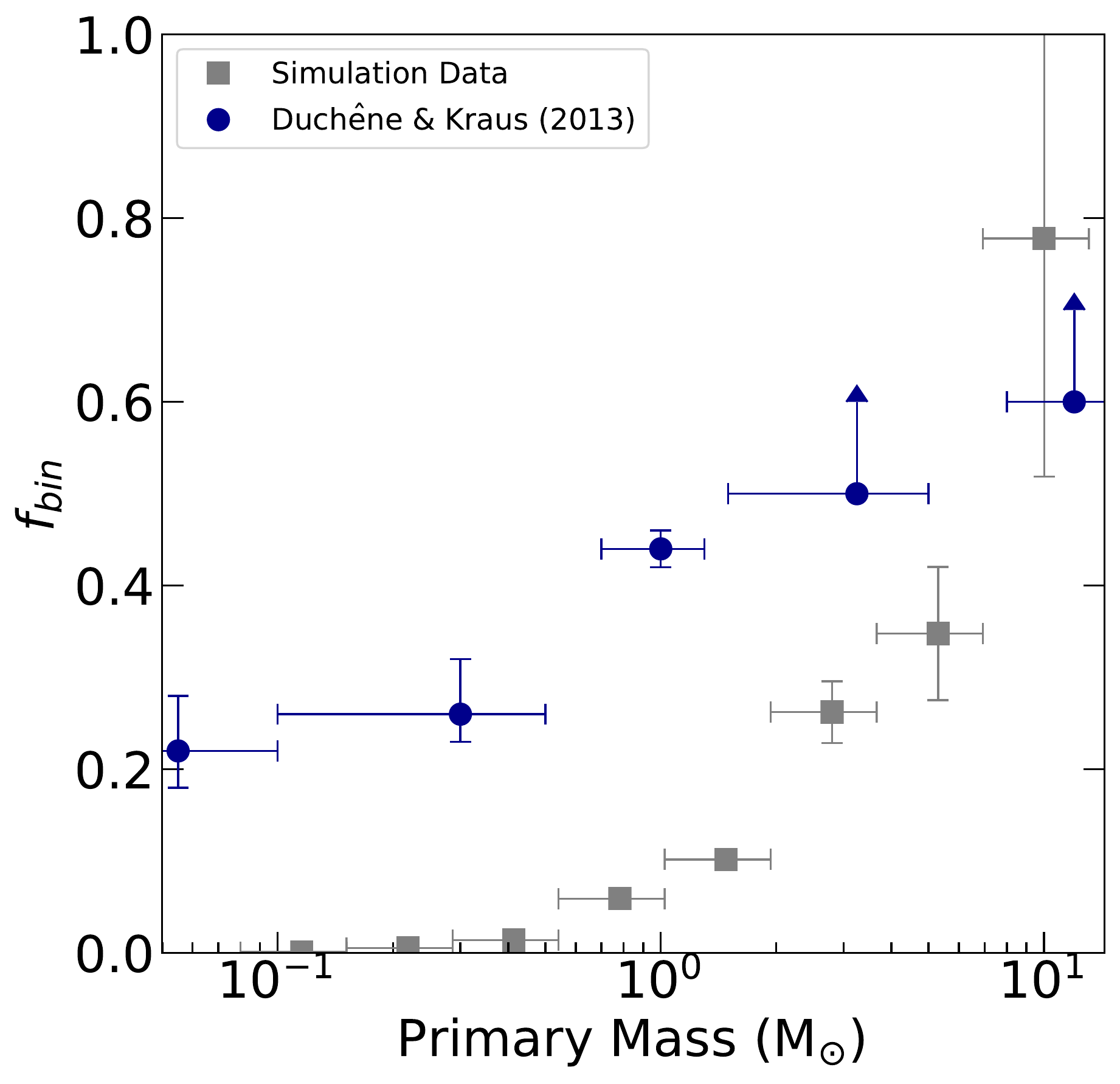}
  \caption{The fraction of all stars in binaries by stellar mass. 
        Our simulations produce massive binaries at a rate consistent
        with observations, but very few low-mass binaries.
         The mass errors shown are 
    the bin widths, while the
    $f_{bin}$ error is given by the Poisson statistical error
           $N_s^{-1/2}$.\label{fig:frac_bin}}
\end{figure}

\begin{figure}[h!]
  \includegraphics[width=\linewidth]{\bpath 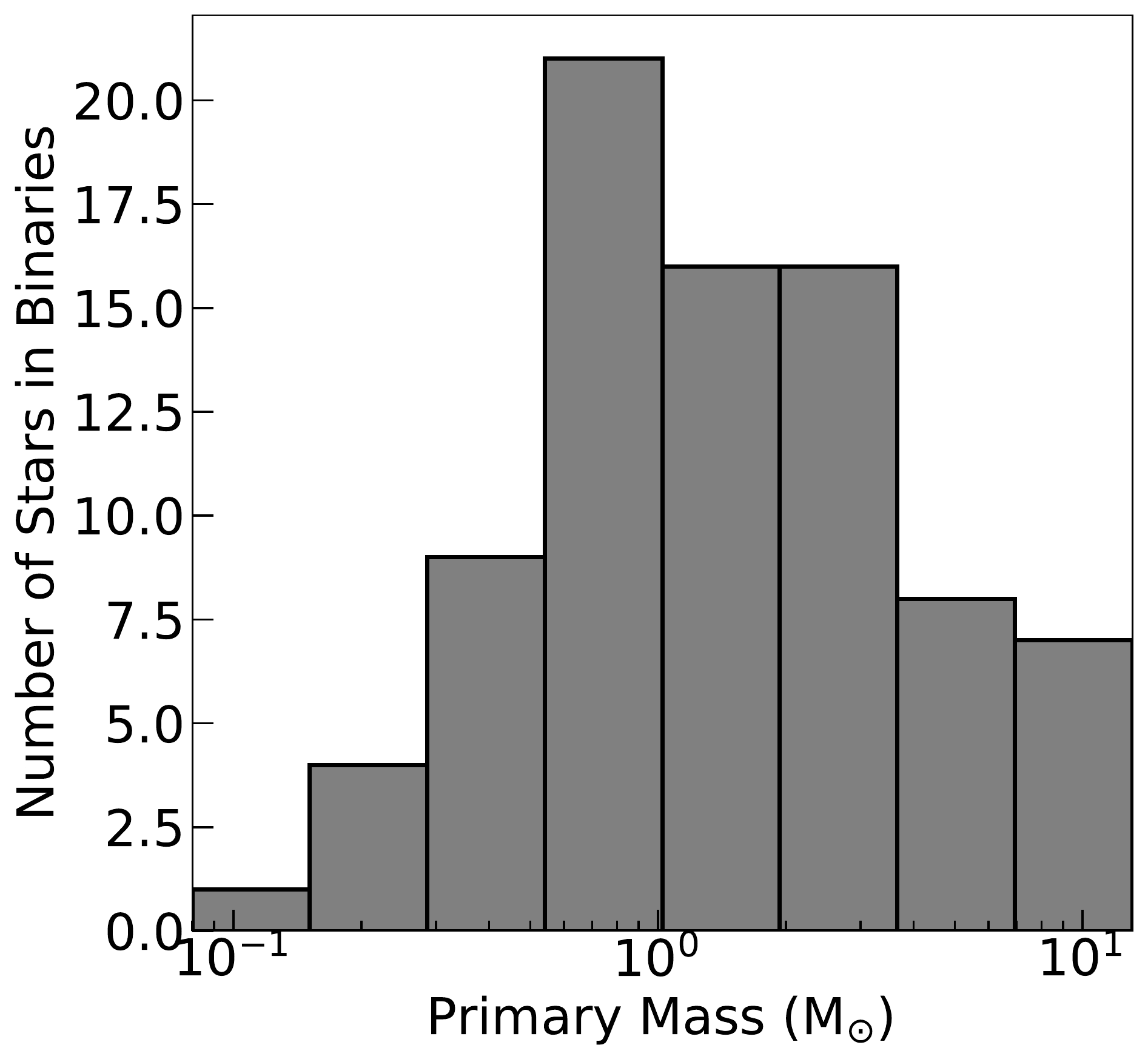}
  \caption{The number of stars in binaries binned by stellar
    mass. \label{fig:num_bin}}
\end{figure}

    The multiplicity fraction
    \begin{equation} f_{bin} = \frac{B}{S + B} \end{equation}
for each mass bin is shown in Fig.\ \ref{fig:frac_bin}, 
    where $B$ is the total number of binary systems, and $S$ is the
    total number of single stars.  
For comparison we include observations of $f_{bin}$ 
  compiled by \citet{Duchene_Kraus_Stellar_Mult_2013ARA&A..51..269D}. 

    The lack of low mass binaries is due to the fact that
we do not include any primordial binaries as we form stars, nor do we
have high enough resolution to capture the gas dynamics that may lead
to primordial binary formation, such as core fragmentation at small scales
\citep{Bate_2012_stellar_multiplicity}. Fig.~\ref{fig:num_bin} 
shows that  in absolute numbers, most binaries are close to 
1 $\msun$, with a steep decline for more massive stars following the IMF.
We have no binaries containing a primary star with mass below $0.1 \msun$,
although our IMF goes down to $0.08 \msun$ in all runs. 

\begin{figure}[h!]
	\includegraphics[width=\linewidth]{\bpath 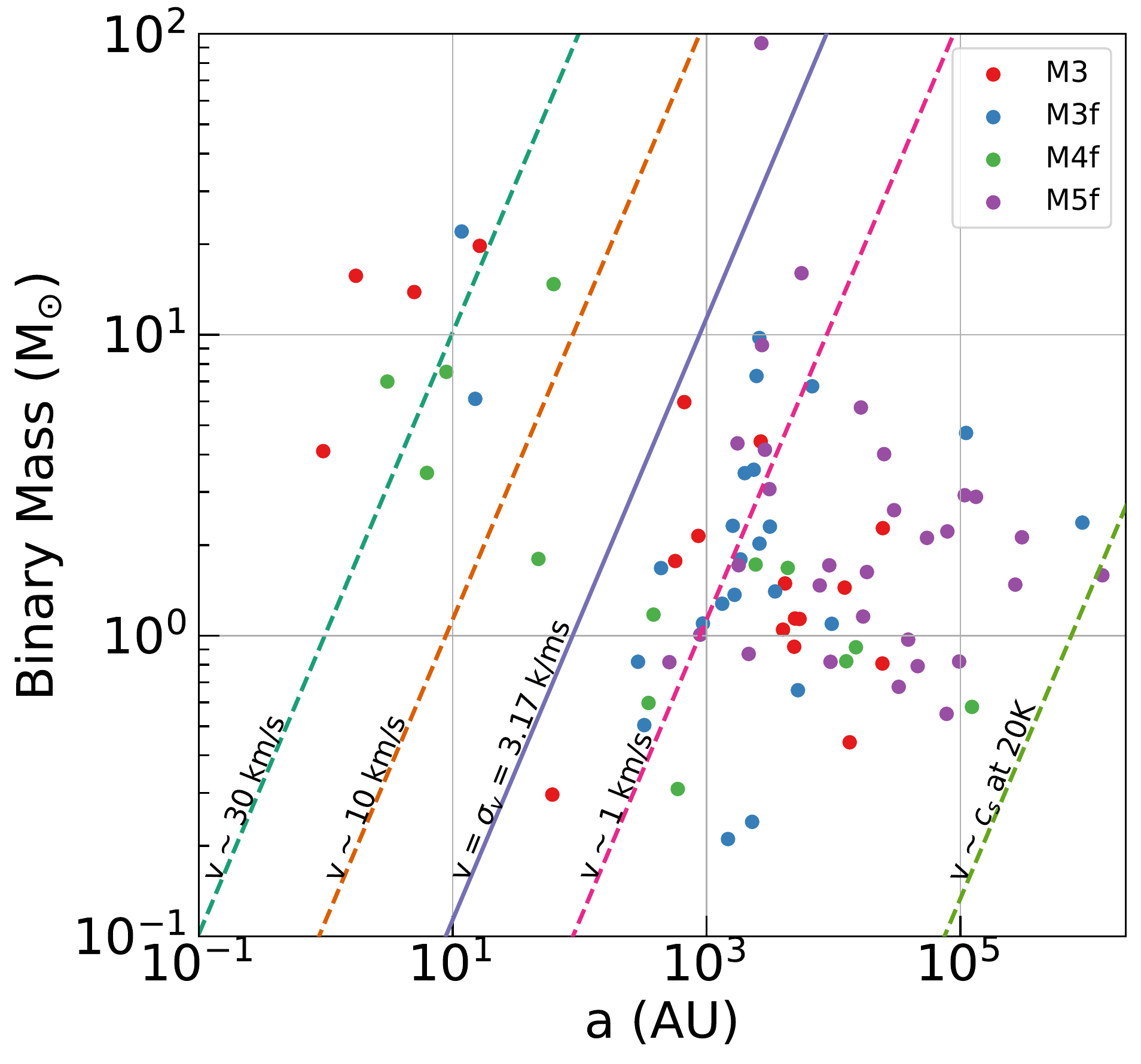}
	\caption{The binary mass and semi-major axis, with lines representing several cluster thermal energies overplotted on their corresponding binary potential masses and radii. Also shown is the mean thermal energy of all the stars averaged across all four simulations. There is a clear separation into hard and soft binaries, as expected from the Heggie-Hills Law.\label{fig:bm_v_a}}
\end{figure}

The value of $f_{bin}$ at the massive end appears remarkably
consistent with the observations, despite our neglect of 
primordial binaries. 

Indeed, all of our massive binaries have separations $r \gtrsim
\SI{1}{AU}$, consistent with dynamical formation,
   as shown in Fig.~\ref{fig:bm_v_a}. 
This agrees with observations that show the majority of massive stars 
occur in hierarchical systems consisting of tight, presumably
primordial, binaries orbiting on wide orbits consistent with the
dynamical binaries formed in our system \citep{gravity_collaboration_multiple_2018}.

\begin{figure}[h!]
  \includegraphics[width=\linewidth]{\bpath 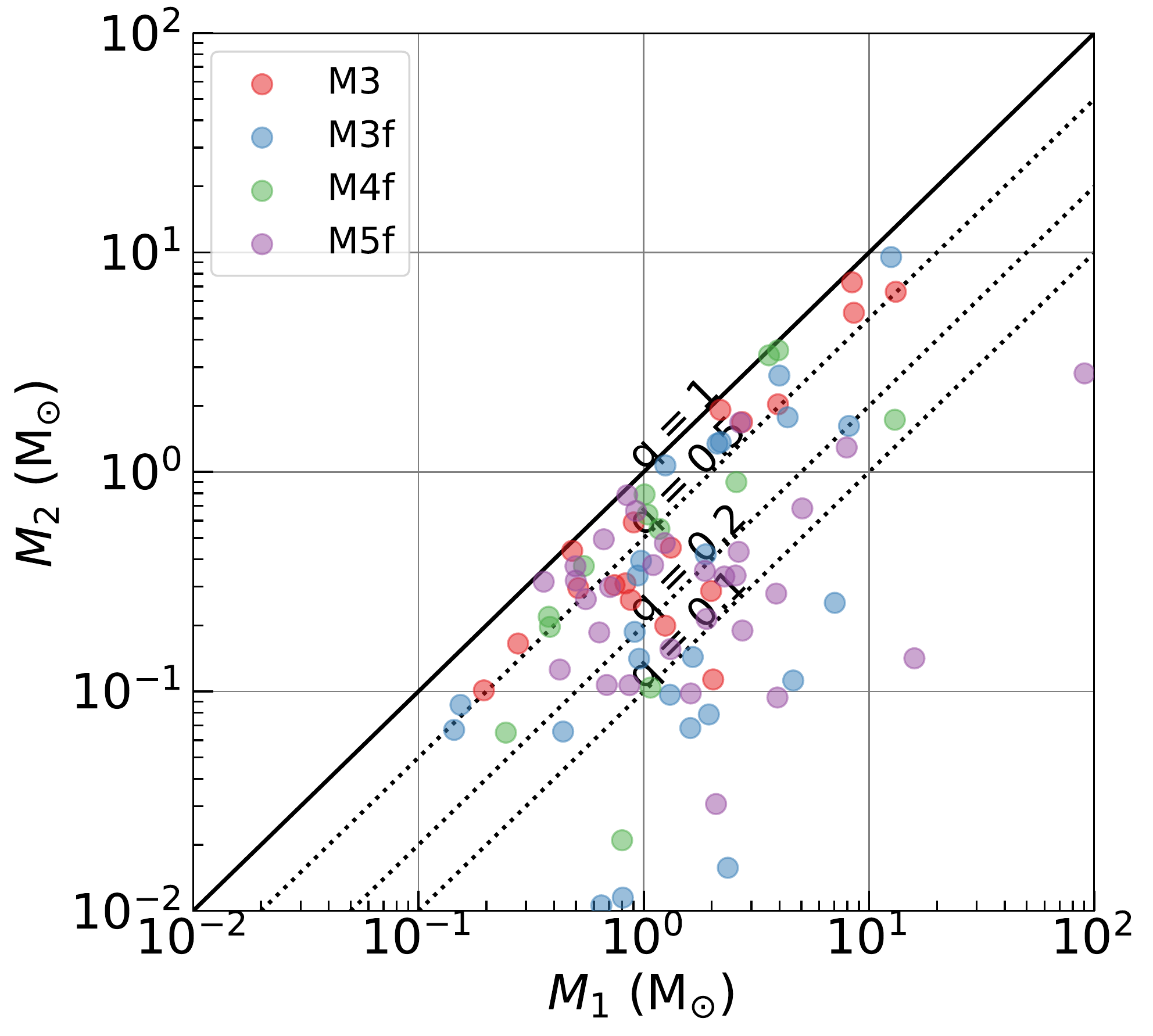}
  \caption{The mass ratio of the primary to the secondary of the 77
    binaries we have found in our runs so far. Also shown are several
    lines of constant mass ratio $q$ for
    comparison. \label{fig:q_bin}}
\end{figure}

\begin{figure}[h!]
  \includegraphics[width=\linewidth]{\bpath 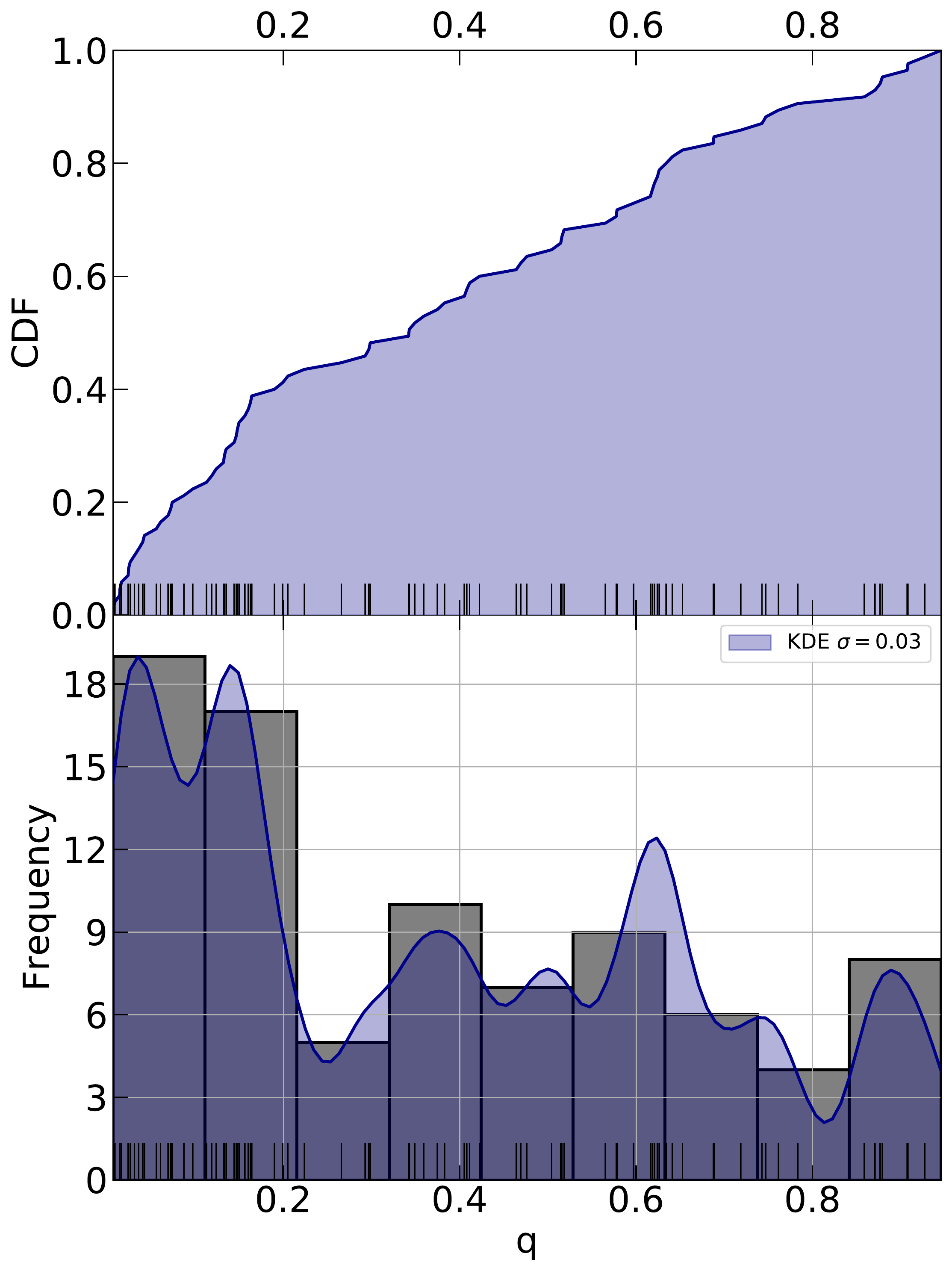}
  \caption{The distribution of mass ratios. In the top panel we show a
    cumulative distribution, while in the bottom panel we show a histogram
    overplotted with a kernel density estimation. The Gaussian kernel
    for this estimate has a
    bandwidth $\sigma=0.03$.  \label{fig:q_hist}}
\end{figure}

For binaries consisting of a primary mass $M_p$ 
and secondary mass $M_s$, the mass ratio $q=M_s / M_p$, is shown in 
Figs.~\ref{fig:q_bin} and
\ref{fig:q_hist}. Comparing with observations in our mass range, we
find our $q$ distribution consistent with \citet{Kouwenhoven_prim_bin_one}, 
with a large peak for $q < 0.2$ followed by a power law drop. It is interesting to 
note that both our data and \citet{Kouwenhoven_prim_bin_one} appear to have moderate 
peaks near $q \sim 0.5$ and $q > 0.8$.
 To test the robustness of the peaks in our histogram, we also
examine the data as a cumulative distribution and using kernel
density estimation.
The multi-modal appearance of the data is evident
in all three methods. The width of the bins for the histogram
was calculated with the method described by \citet{Doane_hist_bins_1976},
which works well for small data sets and does not assume the data is
strictly Gaussian. For the kernel density estimation 
we use the cross validation technique
of leaving out one data point for computing the bandwidth $\sigma$ of
the Gaussian. We evenly sample $\sigma$ from $10^{-3}$ to $1$ for each
rotation through all the data, comparing the mean integrated square
error of each fit with the full data set, to find the appropriate
bandwidth.

With fully collisional N-body dynamics, we expect to see a separation
of our binaries into hard and soft regimes following the Heggie-Hills law 
\citep{Hills_Law_1975,Heggie_binaries_1975MNRAS.173..729H}.
For an average effective cluster thermal energy of
\begin{eqnarray}
  {\textstyle\frac32}NkT = {\textstyle\frac12} \left\langle m \right\rangle \sigma_v^2,
\end{eqnarray}
and a binary energy of
\begin{eqnarray}
x = \frac{G m_1 m_2}{2a},
\end{eqnarray}
with soft binaries having $x/kT \lesssim 1$ and hard binaries having 
$x/kT \gtrsim 1$, the separation between the two types
grows with time. In our runs, we 
indeed see distinct hard and soft populations well separated by a 
boundary at 
$\sigma_v \sim \SI{3}{km~s^{-1}}$, the stellar velocity
dispersion averaged across all four runs. It also appears that the
soft binaries accumulate near the hard/soft boundary, which should be
an important energy sink for the clusters as these binaries are
disrupted. 
Gas dynamical friction could drive binaries to
build up in this way, after which they could be disrupted near the
maximal soft binary energy and thereby drive the entire cluster to
contract \citep{Leigh_GDF_cluster}. Higher resolution runs will be needed to confirm 
if this effect occurs at the smallest binary separations.

\section{Summary}
\label{conclusions}

In this work we have coupled the Eulerian MHD code {\flash} with
stellar evolution and collisional N-body dynamics using the {\phfour} 
and \seba codes in the \amuse\ software framework.
We then used {\amuse} to couple the two gravity calculations
using a gravity bridge to allow for interaction between the gas
and stars, allowing us to simulate open cluster formation and early
evolution in spherical, turbulent clouds of masses $10^3$--$10^5$~$M_{\odot}$.

We have examined the binary populations produced in our demonstration
runs. Despite not injecting any primordial binary population from core or disk
fragmentation, we find a large number of wide binaries with
properties that suggest they formed by interaction with the gas.

We find that the mass ratios of these binaries appear consistent with
observations, and that the binary fraction of massive binaries is close to that
observed.  The lack of low-mass or tight binaries that we find suggests that
those populations are predominantly produced by 
primordial core or disk fragmentation, but that the wide
hierarchical multiple systems in which massive stars occur may be formed by
this dynamical mechanism acting on primordial binaries.
We find well separated hard and soft binary
populations as predicted by the Heggie-Hills law, with evidence of a 
buildup of soft binaries near the boundary between the groups.  
Our results suggest that the hitherto little considered
interaction of stars with gas during the early evolution of
stellar clusters while their natal gas remains present, may explain
much of the wide binary and multiple population, particularly for
the most massive stars.

With publication of this work we 
make public our interface for the {\flash} code in the {\amuse} framework, in order 
to allow reproduction of this work. We hope our interface inspires others to use the 
coupling ideas behind this work in ways we might never consider ourselves, in the 
spirit of scientific discovery.
   The interface can be found within the {\flash} directory of the
   {\amuse} code at \url{https://github.com/amusecode/amuse}. Specific
   implementation details are available from the first author upon
   request.

\acknowledgements
We acknowledge M. Davis, C. Federrath, S. Glover, A. Hill, J. Moreno,
\& E. Pellegrini for useful discussions,
 and A. van Elteran and I. Pelupessy for assistance with AMUSE,
 R. Banerjee and D. Seifried for kindly providing the base code for
 dust and molecular cooling, and R. W\"unsch for kindly providing a
 helper script for the initial conditions. Analysis and plotting for
 this work was done using {\tt yt} \citep{yt}. M-MML was additionally
 supported by the Alexander-von-Humboldt Stiftung.  We acknowledge
 NASA grant NNX14AP27G, which supported this work, and the Dutch
 National Supercomputing Center SURFSara grant 15520 that provided
 computing resources for our simulations.
    RSK acknowledges support from the Deutsche
    Forschungsgemeinschaft (DFG) via SFB 881 "The Milky Way System"
    (sub-projects B1, B2 and B8), and SPP 1573 "Physics of the
    ISM". Furthermore RSK thanks the European Research Council for
    funding under the European Community’s Seventh Framework Programme
    via the ERC Advanced Grant "STARLIGHT" (project number 339177). 

\bibliography{bridge.bbl}

\begin{thebibliography}{62}
\expandafter\ifx\csname natexlab\endcsname\relax\def\natexlab#1{#1}\fi

\bibitem[{{Aarseth} \& {Zare}(1974)}]{Aarseth_KS_regularization_1974}
{Aarseth}, S.~J., \& {Zare}, K. 1974,
  \href{http://dx.doi.org/10.1007/BF01227619}{Celestial Mechanics, 10, 185}

\bibitem[{{Bate}(2009)}]{bate2009}
{Bate}, M.~R. 2009, \mnras, 392, 1363

\bibitem[{{Bate}(2012)}]{Bate_2012_stellar_multiplicity}
---. 2012, \href{http://dx.doi.org/10.1111/j.1365-2966.2011.19955.x}{\mnras,
  419, 3115}

\bibitem[{{Bate} {et~al.}(1995){Bate}, {Bonnell}, \& {Price}}]{Bate_Sinks_1995}
{Bate}, M.~R., {Bonnell}, I.~A., \& {Price}, N.~M. 1995, \mnras, 277, 362

\bibitem[{Binney \& Tremaine(2011)}]{binney2011galactic}
Binney, J., \& Tremaine, S. 2011, Galactic Dynamics: (Second Edition),
  Princeton Series in Astrophysics (Princeton University Press)

\bibitem[{{Chatterjee} {et~al.}(2010){Chatterjee}, {Fregeau}, {Umbreit}, \&
  {Rasio}}]{chatterjee2010}
{Chatterjee}, S., {Fregeau}, J.~M., {Umbreit}, S., \& {Rasio}, F.~A. 2010,
  \href{http://dx.doi.org/10.1088/0004-637X/719/1/915}{\apj, 719, 915}

\bibitem[{{Colella} \& {Woodward}(1984)}]{Colella_and_Woodward_PPM_1984}
{Colella}, P., \& {Woodward}, P.~R. 1984,
  \href{http://dx.doi.org/10.1016/0021-9991(84)90143-8}{J. Comput.\ Phys., 54,
  174}

\bibitem[{{Dale} {et~al.}(2014){Dale}, {Ngoumou}, {Ercolano}, \&
  {Bonnell}}]{Dale_Winds_and_H2}
{Dale}, J.~E., {Ngoumou}, J., {Ercolano}, B., \& {Bonnell}, I.~A. 2014, \mnras,
  442, 694

\bibitem[{Doane(1976)}]{Doane_hist_bins_1976}
Doane, D.~P. 1976, \href{http://dx.doi.org/10.1080/00031305.1976.10479172}{The
  American Statistician, 30, 181}

\bibitem[{{Duch{\^e}ne} \&
  {Kraus}(2013)}]{Duchene_Kraus_Stellar_Mult_2013ARA&A..51..269D}
{Duch{\^e}ne}, G., \& {Kraus}, A. 2013,
  \href{http://dx.doi.org/10.1146/annurev-astro-081710-102602}{\araa, 51, 269}

\bibitem[{Elmegreen(1997)}]{elmegreen_IMF_Poisson_1997}
Elmegreen, B.~G. 1997, \href{http://dx.doi.org/10.1086/304562}{\apj, 486, 944}

\bibitem[{{Federrath} {et~al.}(2010){Federrath}, {Banerjee}, {Clark}, \&
  {Klessen}}]{Federrath_Sink_Particles}
{Federrath}, C., {Banerjee}, R., {Clark}, P.~C., \& {Klessen}, R.~S. 2010,
  \apj, 713, 269

\bibitem[{Federrath {et~al.}(2011)Federrath, Sur, Schleicher, Banerjee, \&
  Klessen}]{federrath_new_jeans_2011}
Federrath, C., Sur, S., Schleicher, D. R.~G., Banerjee, R., \& Klessen, R.~S.
  2011, \href{http://dx.doi.org/10.1088/0004-637X/731/1/62}{\apj, 731, 62}

\bibitem[{{Fryxell} {et~al.}(2000){Fryxell}, {Olson}, {Ricker}, {Timmes},
  {Zingale}, {Lamb}, {MacNeice}, {Rosner}, {Truran}, \& {Tufo}}]{FLASH}
{Fryxell}, B., {Olson}, K., {Ricker}, P., {et~al.} 2000,
  \href{http://dx.doi.org/10.1086/317361}{\apjs, 131, 273}

\bibitem[{{Fujii} {et~al.}(2007){Fujii}, {Iwasawa}, {Funato}, \&
  {Makino}}]{Fujii_Makino_bridge}
{Fujii}, M., {Iwasawa}, M., {Funato}, Y., \& {Makino}, J. 2007,
  \href{http://dx.doi.org/10.1093/pasj/59.6.1095}{\pasj, 59, 1095}

\bibitem[{Fujii \& {Portegies Zwart}(2015)}]{fujii_zwart_IMF_2015}
Fujii, M.~S., \& {Portegies Zwart}, S. 2015,
  \href{http://dx.doi.org/10.1093/mnras/stv293}{\mnras, 449, 726}

\bibitem[{{Gatto} {et~al.}(2017){Gatto}, {Walch}, {Naab}, {Girichidis},
  {W{\"u}nsch}, {Glover}, {Klessen}, {Clark}, {Peters}, {Derigs}, {Baczynski},
  \& {Puls}}]{Gatto_Walch_SILCC3_2016}
{Gatto}, A., {Walch}, S., {Naab}, T., {et~al.} 2017,
  \href{http://dx.doi.org/10.1093/mnras/stw3209}{\mnras, 466, 1903}

\bibitem[{{Heggie}(1975)}]{Heggie_binaries_1975MNRAS.173..729H}
{Heggie}, D.~C. 1975, \href{http://dx.doi.org/10.1093/mnras/173.3.729}{\mnras,
  173, 729}

\bibitem[{Heitsch {et~al.}(2001)Heitsch, Mac~Low, \&
  Klessen}]{heitsch_gravitational_2001}
Heitsch, F., Mac~Low, M.-M., \& Klessen, R.~S. 2001,
  \href{http://dx.doi.org/10.1086/318335}{\apj, 547, 280}

\bibitem[{{Hills}(1975)}]{Hills_Law_1975}
{Hills}, J.~G. 1975, \href{http://dx.doi.org/10.1086/111815}{\aj, 80, 809}

\bibitem[{{Hut} {et~al.}(1995){Hut}, {Makino}, \& {McMillan}}]{Hut_better_LF}
{Hut}, P., {Makino}, J., \& {McMillan}, S. 1995,
  \href{http://dx.doi.org/10.1086/187844}{\apj, 443, L93}

\bibitem[{{Hypki} \& {Giersz}(2013)}]{hypki2013}
{Hypki}, A., \& {Giersz}, M. 2013,
  \href{http://dx.doi.org/10.1093/mnras/sts415}{\mnras, 429, 1221}

\bibitem[{{Ib{\'a}{\~n}ez-Mej{\'{\i}}a}
  {et~al.}(2016){Ib{\'a}{\~n}ez-Mej{\'{\i}}a}, {Mac Low}, {Klessen}, \&
  {Baczynski}}]{ibanez-mejia_gravitational_2015}
{Ib{\'a}{\~n}ez-Mej{\'{\i}}a}, J.~C., {Mac Low}, M.-M., {Klessen}, R.~S., \&
  {Baczynski}, C. 2016, \apj, 824, 41

\bibitem[{{Joung} \& {Mac Low}(2006)}]{Joung_SN_driven_turb}
{Joung}, M.~K.~R., \& {Mac Low}, M.-M. 2006,
  \href{http://dx.doi.org/10.1086/508795}{\apj, 653, 1266}

\bibitem[{Karl {et~al.}(2018)Karl, Pfuhl, Eisenhauer, Genzel, Grellmann,
  Habibi, Abuter, Accardo, Amorim, Anugu, Ávila, Benisty, Berger, Bland,
  Bonnet, Bourget, Brandner, Brast, Buron, Garatti, Chapron, Clénet, Collin,
  Foresto, de~Wit, de~Zeeuw, Deen, Delplancke-Ströbele, Dembet, Derie, Dexter,
  Duvert, Ebert, Eckart, Esselborn, Fédou, Finger, Garcia, Dabo, Lopez, Gao,
  Gandron, Gillessen, Gonté, Gordo, Grözinger, Guajardo, Guieu, Haguenauer,
  Hans, Haubois, Haug, Haußmann, Henning, Hippler, Horrobin, Huber, Hubert,
  Hubin, Hummel, Jakob, Jochum, Jocou, Kaufer, Kellner, Kandrew, Kern,
  Kervella, Kiekebusch, Klein, Köhler, Kolb, Kulas, Lacour, Lapeyrère,
  Lazareff, Bouquin, Léna, Lenzen, Lévêque, Lin, Lippa, Magnard, Mehrgan,
  Mérand, Moulin, Müller, Müller, Neumann, Oberti, Ott, Pallanca, Panduro,
  Pasquini, Paumard, Percheron, Perraut, Perrin, Pflüger, Duc, Plewa, Popovic,
  Rabien, Ramírez, Ramos, Rau, Riquelme, Rodríguez-Coira, Rohloff, Rosales,
  Rousset, Sanchez-Bermudez, Scheithauer, Schöller, Schuhler, Spyromilio,
  Straub, Straubmeier, Sturm, Suarez, Tristram, Ventura, Vincent, Waisberg,
  Wank, Widmann, Wieprecht, Wiest, Wiezorrek, Wittkowski, Woillez, Wolff,
  Yazici, Ziegler, \& Zins}]{gravity_collaboration_multiple_2018}
Karl, M., Pfuhl, O., Eisenhauer, F., {et~al.} 2018,
  \href{http://arxiv.org/abs/1809.10376}{arXiv:1809.10376 [astro-ph]}

\bibitem[{{Kouwenhoven} {et~al.}(2005){Kouwenhoven}, {Brown}, {Zinnecker},
  {Kaper}, \& {Portegies Zwart}}]{Kouwenhoven_prim_bin_one}
{Kouwenhoven}, M.~B.~N., {Brown}, A.~G.~A., {Zinnecker}, H., {Kaper}, L., \&
  {Portegies Zwart}, S.~F. 2005,
  \href{http://dx.doi.org/10.1051/0004-6361:20048124}{\aap, 430, 137}

\bibitem[{Kroupa(2001)}]{kroupa_IMF_2001}
Kroupa, P. 2001,
  \href{http://dx.doi.org/10.1046/j.1365-8711.2001.04022.x}{\mnras, 322, 231}

\bibitem[{{Krumholz} {et~al.}(2007){Krumholz}, {Klein}, \&
  {McKee}}]{krumholz2007}
{Krumholz}, M.~R., {Klein}, R.~I., \& {McKee}, C.~F. 2007, \apj, 656, 959

\bibitem[{{Krumholz} {et~al.}(2004){Krumholz}, {McKee}, \&
  {Klein}}]{Krumholz_Sinks}
{Krumholz}, M.~R., {McKee}, C.~F., \& {Klein}, R.~I. 2004,
  \href{http://dx.doi.org/10.1086/421935}{\apj, 611, 399}

\bibitem[{{Lee}(2013)}]{Lee_USM_2013}
{Lee}, D. 2013, \href{http://dx.doi.org/10.1016/j.jcp.2013.02.049}{J. Comput.\
  Phys., 243, 269}

\bibitem[{{Leigh} {et~al.}(2014){Leigh}, {Mastrobuono-Battisti}, {Perets}, \&
  {B{\"o}ker}}]{Leigh_GDF_cluster}
{Leigh}, N. W.~C., {Mastrobuono-Battisti}, A., {Perets}, H.~B., \& {B{\"o}ker},
  T. 2014, \href{http://dx.doi.org/10.1093/mnras/stu622}{\mnras, 441, 919}

\bibitem[{{Mac Low} \& {Klessen}(2004)}]{Mac_Low_Star_Formation}
{Mac Low}, M.-M., \& {Klessen}, R.~S. 2004, \rmp, 76, 125

\bibitem[{Mac~Low {et~al.}(1998)Mac~Low, Klessen, Burkert, \&
  Smith}]{Mac_Low_Kinetic_1998}
Mac~Low, M.-M., Klessen, R.~S., Burkert, A., \& Smith, M.~D. 1998,
  \href{http://dx.doi.org/10.1103/PhysRevLett.80.2754}{\prl, 80, 2754}

\bibitem[{{Mardling}(2008)}]{mardling2008dynamical}
{Mardling}, R.~A. 2008, \href{http://dx.doi.org/10.1017/S1743921308015615}{in
  IAU Symposium, Vol. 246, Dynamical Evolution of Dense Stellar Systems, ed.
  E.~{Vesperini}, M.~{Giersz}, \& A.~{Sills}}, 199

\bibitem[{{Matzner} \& {Jumper}(2015)}]{matzner2015}
{Matzner}, C.~D., \& {Jumper}, P.~H. 2015, \apj, 815, 68

\bibitem[{{McMillan} \& {Hut}(1996)}]{McMillan_Hut_smallN}
{McMillan}, S. L.~W., \& {Hut}, P. 1996,
  \href{http://dx.doi.org/10.1086/177610}{\apj, 467, 348}

\bibitem[{{McMillan, S.} {et~al.}(2012){McMillan, S.}, {van Elteran, A.}, \&
  {Whitehead, A.}}]{ph4}
{McMillan, S.}, {van Elteran, A.}, \& {Whitehead, A.} 2012, in {Astronomical
  Society of the Pacific Conference Series}, Vol. 453, 129

\bibitem[{{Miyoshi} \& {Kusano}(2005)}]{Miyoshi_HLLD_solver}
{Miyoshi}, T., \& {Kusano}, K. 2005,
  \href{http://dx.doi.org/10.1016/j.jcp.2005.02.017}{J. Comput.\ Phys., 208,
  315}

\bibitem[{Neufeld {et~al.}(1995)Neufeld, Lepp, \&
  Melnick}]{neufeld_thermal_1995}
Neufeld, D.~A., Lepp, S., \& Melnick, G.~J. 1995,
  \href{http://dx.doi.org/10.1086/192211}{\apjs, 100, 132}

\bibitem[{Padoan {et~al.}(2014)Padoan, Haugb{\o}lle, \&
  Nordlund}]{padoan_infall-driven_2014}
Padoan, P., Haugb{\o}lle, T., \& Nordlund, {\AA}. 2014,
  \href{http://dx.doi.org/10.1088/0004-637X/797/1/32}{\apj, 797, 32}

\bibitem[{{Paxton} {et~al.}(2011){Paxton}, {Bildsten}, {Dotter}, {Herwig},
  {Lesaffre}, \& {Timmes}}]{MESA_2011ApJS..192....3P}
{Paxton}, B., {Bildsten}, L., {Dotter}, A., {et~al.} 2011,
  \href{http://dx.doi.org/10.1088/0067-0049/192/1/3}{\apjs, 192, 3}

\bibitem[{{Pelupessy} \& {Portegies Zwart}(2012)}]{Pelupessy_embedded_SC}
{Pelupessy}, F.~I., \& {Portegies Zwart}, S. 2012, \mnras, 420, 1503

\bibitem[{{Pelupessy} {et~al.}(2013){Pelupessy}, {van Elteren}, {de Vries},
  {McMillan}, {Drost}, \& {Portegies Zwart}}]{AMUSE}
{Pelupessy}, F.~I., {van Elteren}, A., {de Vries}, N., {et~al.} 2013,
  \href{http://dx.doi.org/10.1051/0004-6361/201321252}{\aap, 557, A84}

\bibitem[{{Peters} {et~al.}(2011){Peters}, {Banerjee}, {Klessen}, \& {Mac
  Low}}]{peters2011}
{Peters}, T., {Banerjee}, R., {Klessen}, R.~S., \& {Mac Low}, M.-M. 2011,
  \href{http://dx.doi.org/10.1088/0004-637X/729/1/72}{\apj, 729, 72}

\bibitem[{{Peters} {et~al.}(2010){Peters}, {Klessen}, {Mac Low}, \&
  {Banerjee}}]{peters2010b}
{Peters}, T., {Klessen}, R.~S., {Mac Low}, M.-M., \& {Banerjee}, R. 2010,
  \href{http://dx.doi.org/10.1088/0004-637X/725/1/134}{\apj, 725, 134}

\bibitem[{{Portegies Zwart} {et~al.}(2013){Portegies Zwart}, McMillan, van
  Elteren, Pelupessy, \& de~Vries}]{zwart_multi-physics_2013}
{Portegies Zwart}, S., McMillan, S., van Elteren, A., Pelupessy, I., \&
  de~Vries, N. 2013, Comput.\ Phys.\ Comm., 184, 456

\bibitem[{{Portegies Zwart} \& {McMillan}(2019)}]{Mcmillan_Art_of_AMUSE}
{Portegies Zwart}, S., \& {McMillan}, S. L.~W. 2019, Astrophysical Recipes: The
  Art of Amuse (Institute of Physics Publishing)

\bibitem[{{Portegies Zwart} {et~al.}(1999){Portegies Zwart}, {Makino},
  {McMillan}, \& {Hut}}]{Portegies_Zwart_runaway_colls_clusters_1999}
{Portegies Zwart}, S.~F., {Makino}, J., {McMillan}, S.~L.~W., \& {Hut}, P.
  1999, \aap, 348, 117

\bibitem[{{Portegies Zwart} \& {Verbunt}(1996)}]{Portegies_SeBa}
{Portegies Zwart}, S.~F., \& {Verbunt}, F. 1996, \aap, 309, 179

\bibitem[{{Ricker}(2008)}]{Ricker_MG_solver_2008}
{Ricker}, P.~M. 2008, \href{http://dx.doi.org/10.1086/526425}{\apjs, 176, 293}

\bibitem[{{Rosen} {et~al.}(2016){Rosen}, {Krumholz}, {McKee}, \&
  {Klein}}]{rosen_unstable_2016}
{Rosen}, A.~L., {Krumholz}, M.~R., {McKee}, C.~F., \& {Klein}, R.~I. 2016,
  \href{http://dx.doi.org/10.1093/mnras/stw2153}{\mnras, 463, 2553}

\bibitem[{Sadavoy \& Stahler(2017)}]{sadavoy_stahler_embedded_2017}
Sadavoy, S.~I., \& Stahler, S.~W. 2017,
  \href{http://dx.doi.org/10.1093/mnras/stx1061}{\mnras, 469, 3881}

\bibitem[{{Seifried} {et~al.}(2011){Seifried}, {Banerjee}, {Klessen}, {Duffin},
  \& {Pudritz}}]{Seifried_2011MNRAS.417.1054S}
{Seifried}, D., {Banerjee}, R., {Klessen}, R.~S., {Duffin}, D., \& {Pudritz},
  R.~E. 2011, \href{http://dx.doi.org/10.1111/j.1365-2966.2011.19320.x}{\mnras,
  417, 1054}

\bibitem[{{Sormani} {et~al.}(2017){Sormani}, {Tre{\ss}}, {Klessen}, \&
  {Glover}}]{Sormani_sinks_particles}
{Sormani}, M.~C., {Tre{\ss}}, R.~G., {Klessen}, R.~S., \& {Glover}, S. C.~O.
  2017, \href{http://dx.doi.org/10.1093/mnras/stw3205}{\mnras, 466, 407}

\bibitem[{Stahler \& Palla(2008)}]{stahler2008formation}
Stahler, S., \& Palla, F. 2008, The Formation of Stars (Wiley)

\bibitem[{Truelove {et~al.}(1997)Truelove, Klein, McKee, Holliman, Howell, \&
  Greenough}]{truelove_jeans_1997}
Truelove, J.~K., Klein, R.~I., McKee, C.~F., {et~al.} 1997,
  \href{http://dx.doi.org/10.1086/310975}{\apj Letters, 489, L179}

\bibitem[{{Turk} {et~al.}(2011){Turk}, {Smith}, {Oishi}, {Skory}, {Skillman},
  {Abel}, \& {Norman}}]{yt}
{Turk}, M.~J., {Smith}, B.~D., {Oishi}, J.~S., {et~al.} 2011,
  \href{http://dx.doi.org/10.1088/0067-0049/192/1/9}{\apjs, 192, 9}

\bibitem[{Weingartner \& Draine(2001)}]{weingartner_photoelectric_2001}
Weingartner, J.~C., \& Draine, B.~T. 2001,
  \href{http://dx.doi.org/10.1086/320852}{\apjs, 134, 263}

\bibitem[{{Wisdom} \& {Holman}(1991)}]{Wisdom_Holman_1991}
{Wisdom}, J., \& {Holman}, M. 1991, \aj, 102, 1528

\bibitem[{{Wolfire} {et~al.}(2003){Wolfire}, {McKee}, {Hollenbach}, \&
  {Tielens}}]{wolfire_2003}
{Wolfire}, M.~G., {McKee}, C.~F., {Hollenbach}, D., \& {Tielens}, A.~G.~G.~M.
  2003, \href{http://dx.doi.org/10.1086/368016}{\apj, 587, 278}

\bibitem[{W\"{u}nsch(2015)}]{WunschCloud}
W\"{u}nsch, R. 2015, personal communication

\bibitem[{Yoshida(1990)}]{yoshida_construction_1990}
Yoshida, H. 1990, \href{http://dx.doi.org/10.1016/0375-9601(90)90092-3}{Physics
  Letters A, 150, 262}

\end{thebibliography}

\end{document}